\documentclass[useAMS,usenatbib,usegraphicx,longabstract]{mn2e}

\usepackage{times}
\usepackage{epsfig}
\usepackage{epstopdf}

\newcommand{\sauron}{{\texttt {SAURON}}}

\newcommand{\atlas}{ATLAS$^{\rm 3D}$}

\newcommand{\ROSAT}{{\it ROSAT\/}}
\newcommand{\Einstein}{{\it Einstein\/}}
\newcommand{\Chandra}{{\it Chandra\/}}
\newcommand{\XMM}{{\it XMM\/}}

\newcommand{\placefigLBLX}{
\begin{figure}
\begin{center}
\includegraphics[width=\columnwidth]{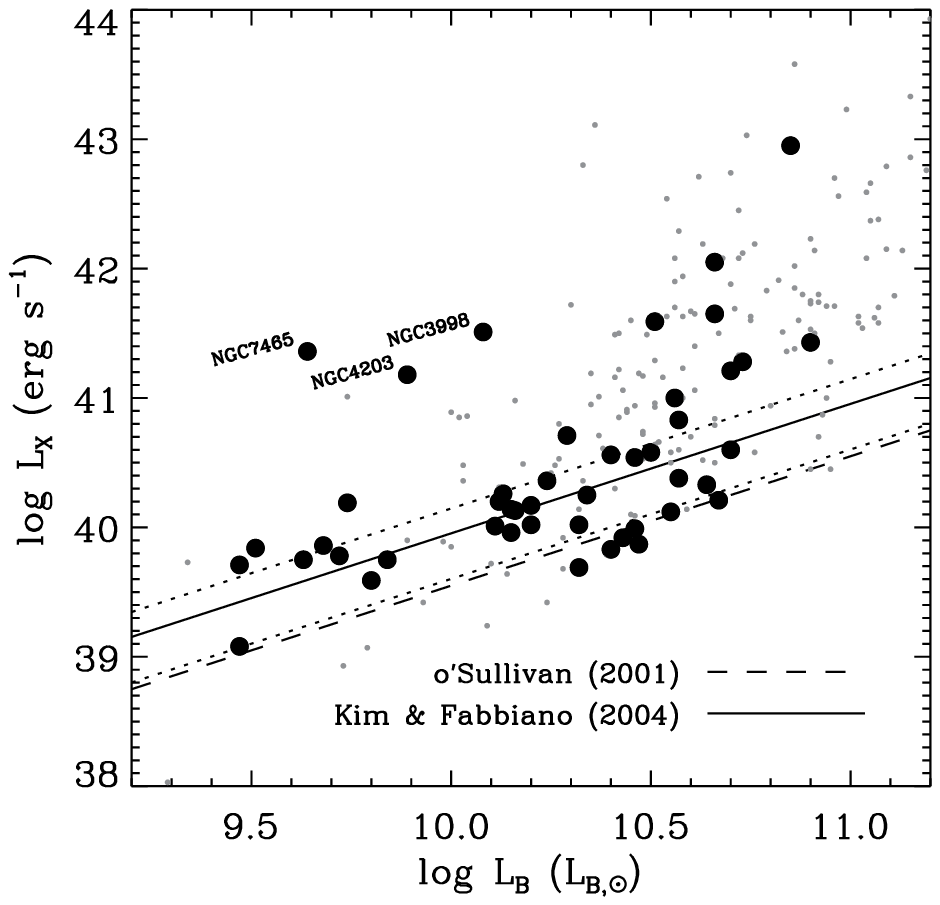}
\end{center}
\caption[]{$L_B$ vs. $L_X$ of our low X-ray resolution ATLAS$^{\rm
    3D}$ sub-sample galaxies, based on luminosity and distance values
  from the literature and essentially (except for one object) from
  \citet{oSu01}.  The dashed and solid lines trace the expected
  contribution to the observed $L_X$ from the unresolved emission of
  low-mass X-ray binaries, as estimated by O'Sullivan et al. and
  \citet{Kim04}, respectively. The dotted lines show the uncertainties
  associated with the \citeauthor{Kim04} calibration.
  The three labelled sources, already identified by O'Sullivan et al.,
  owe most of their X-ray flux to the presence of strong nuclear
  activity. The small grey dots show the other early-type galaxies
  with detected $L_X$ values in the catalogue of O'Sullivan et al.}
\label{fig:LBLX}
\end{figure}
}
\newcommand{\placefigLKLX}{
\begin{figure}
\begin{center}
\includegraphics[width=\columnwidth]{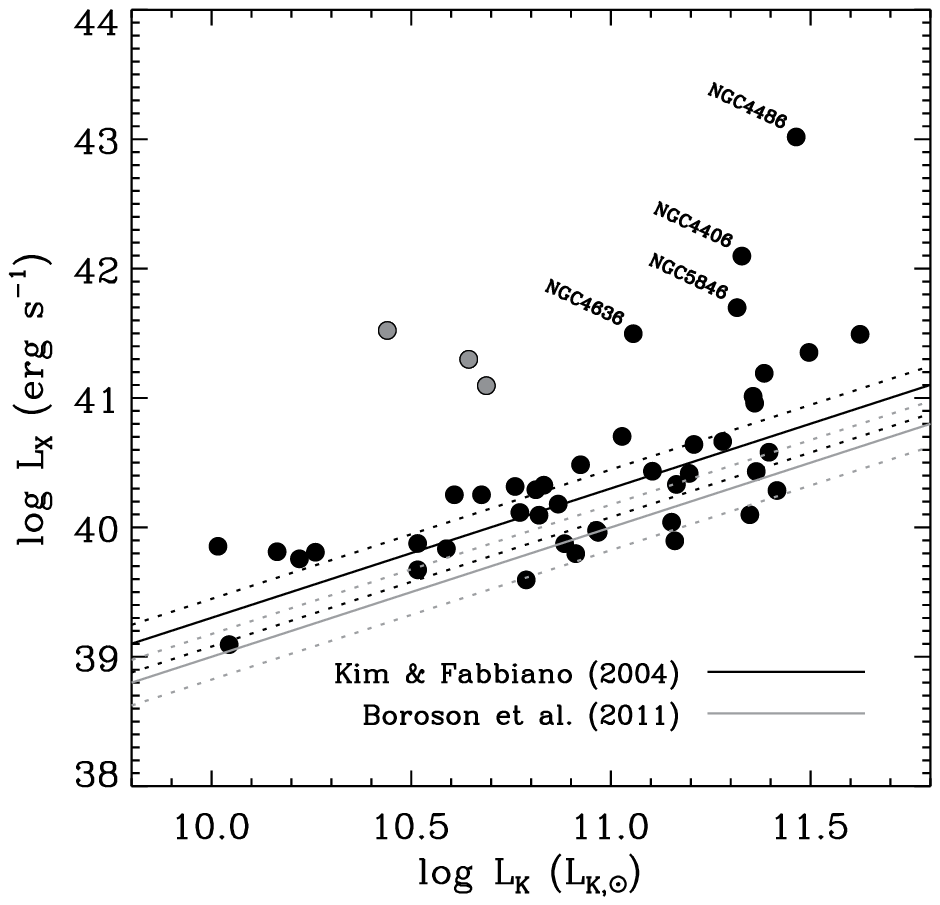}
\end{center}
\caption[]{$L_K$ vs. $L_X$ of our low X-ray resolution ATLAS$^{\rm
    3D}$ sub-sample galaxies, now based on luminosity values obtained
  adopting our own compilation of distance values. The solid and
  dotted lines trace the expected contribution to the observed $L_X$
  values, with uncertainties, from the unresolved emission of low-mass
  X-ray binaries, as given by \citet{Kim04} in black and by
  \citet{Bor11} in grey.
  In this diagram, the labelled objects with the largest X-ray
  luminosities are either deeply embedded in the intracluster medium
  (ICM) of the Virgo cluster (NGC~4486, NGC4406) or are the central
  member of their own group of galaxies and show a rather extended
  X-ray halo (NGC~4636, NGC~5846). Grey symbols show the objects with
  AGNs that were identified in Fig.~\ref{fig:LBLX}.}
\label{fig:LKLX}
\end{figure}
}
\newcommand{\placefigLKsigLX}{
\begin{figure}
\begin{center}
\includegraphics[width=\columnwidth]{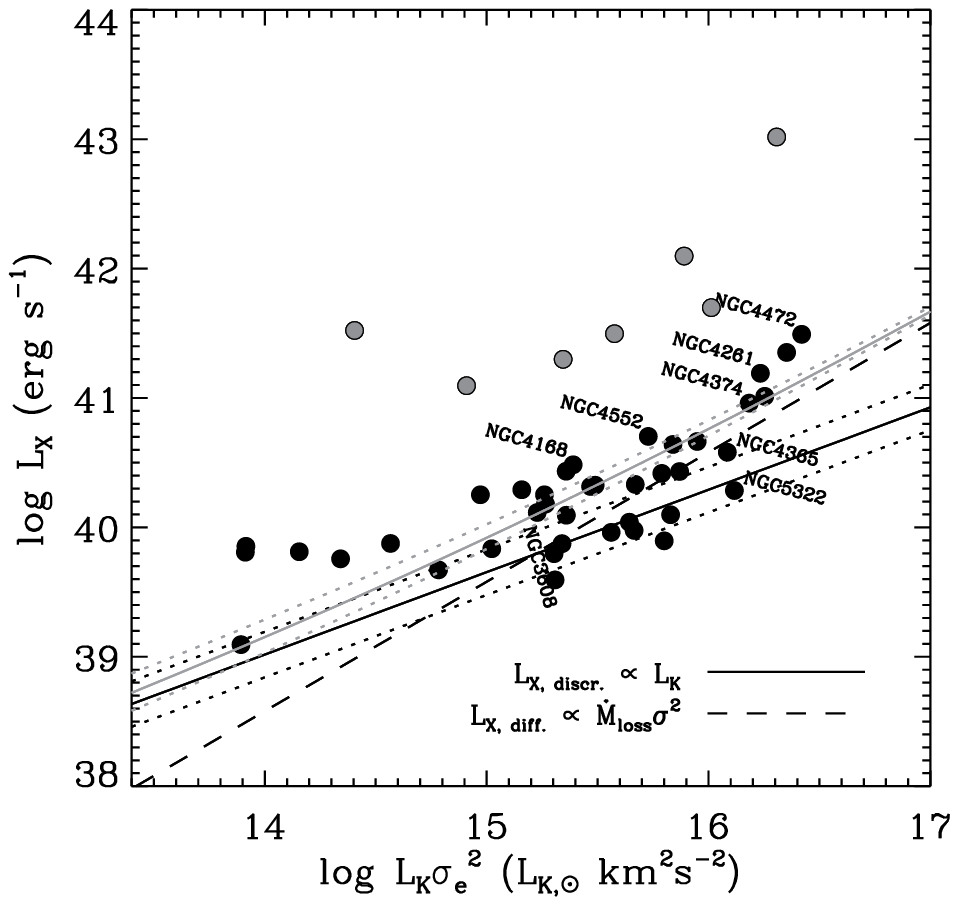}
\end{center}
\caption[]{$L_K\sigma_{\rm e}^2$ vs. $L_X$ diagram for our low X-ray
  resolution ATLAS$^{\rm 3D}$ sub-sample galaxies, to check whether
  their diffuse hot-gas luminosity can be explained in terms of
  stellar-mass loss material that through shocks and collisions
  thermalised the kinetic energy inherited by their parent stars. The
  prediction of such a model, whereby $L_{X,{\rm diff}} = L_{\sigma} =
  \frac{3}{2}\dot M \sigma_{\rm e}^2$, is shown by the dashed line,
  whereas the contribution (with uncertainties) of unresolved X-ray
  binaries \citep[$L_{X,{\rm discr}}$, according to][]{Bor11} is
  shown by the solid and dotted lines (see text for more details on
  both components). Finally, the grey solid and dashed lines show the
  sum of both diffuse and discrete components, which is what we expect
  to observe in the absence of strong AGN or ICM contamination. Beside
  two exceptions, all the slowly-rotating galaxies in our low X-ray
  resolution sample that are massive enough to heat up the
  stellar-loss material to X-ray emitting temperatures (labelled
  objects) lie directly on top of above the prediction of such a
  simple model, whereas most fast rotators fall short of this
  benchmark. Grey symbols indicate either objects with AGNs or that
  are deeply embedded in their group or cluster medium, as shown in
  Figs.~\ref{fig:LBLX} and \ref{fig:LKLX}.}
\label{fig:LKsigLX}
\end{figure}
}
\newcommand{\placefigResidCorrLX}{
\begin{figure*}
\begin{center}
\includegraphics[height=0.95\textwidth,angle=90]{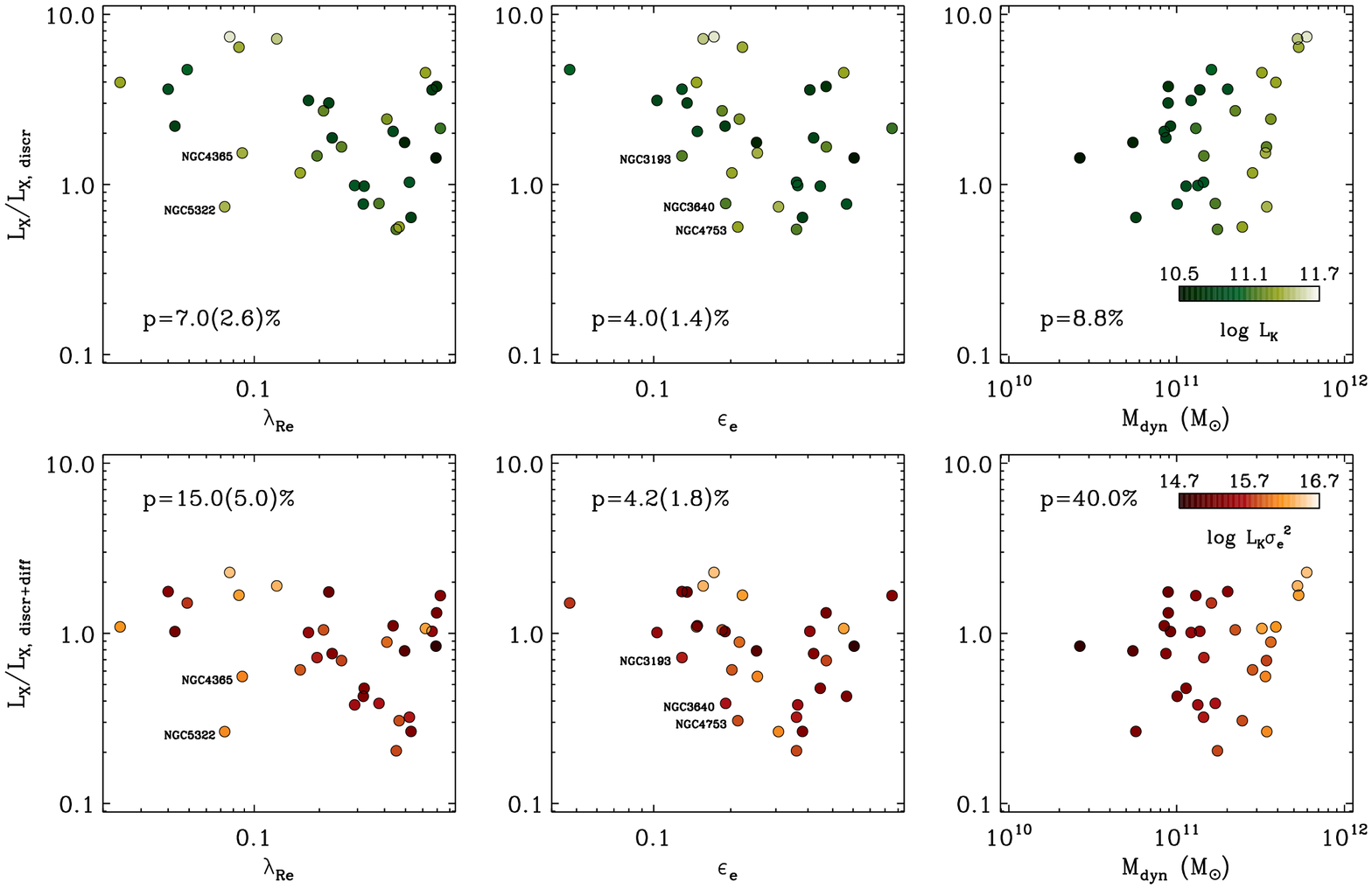}
\end{center}
\caption[]{Correlations between the observed specific angular momentum
  (quantified by the $\lambda_{\rm Re}$ parameter, left panels), the
  apparent flattening ($\epsilon_{\rm e}$, middle panels) and the
  dynamical mass ($M_{\rm dyn}$, right panels) of our low X-ray
  resolution \atlas\ sub-sample galaxies and the ratio of their total
  X-ray luminosity $L_X$ to the expected contribution from unresolved
  discrete sources only ($L_{X,{\rm discr}}$, upper row) or to the
  predicted X-ray luminosity when including also the emission from
  stellar-mass loss material heated to the kinetic temperature of the
  stars ($L_{X,{\rm discr+diff}}$, lower row). In the top and lower
  panels the symbols are colour-coded according to the total K-band
  luminosity $L_K$ and the value of $L_K\sigma_e^2$ for our sample
  galaxies, respectively, in order to trace their position in
  Figs.~\ref{fig:LKLX} and \ref{fig:LKsigLX}. Objects with
  $\log{L_K\sigma_{\rm e}^2} \le 14.6 \,\rm
  L_{K,\odot}\,km^2\,s^{-2}$, which would not be sufficiently massive
  to sustain a detectable X-ray halo against the LMXBs background, and
  NGC~3226, whose $L_X$ value is contaminated by the presence of the
  nearby active galaxy NGC~3227, are not shown. Excluding these
  galaxies, each panel shows the values of the Spearman test
  probability $p$ of the null hypothesis whereby the plotted
  quantities would follow each other monotonically (not necessarily
  linearly) only by mere chance. Values within parentheses refer to
  $p$ values computed while excluding the labelled objects in the
  corresponsing panels.}
\label{fig:ResidCorrLX}
\end{figure*}
}
\newcommand{\placefigLKLXandLKsigLXcol}{
\begin{figure*}
\begin{center}
\includegraphics[width=0.95\textwidth]{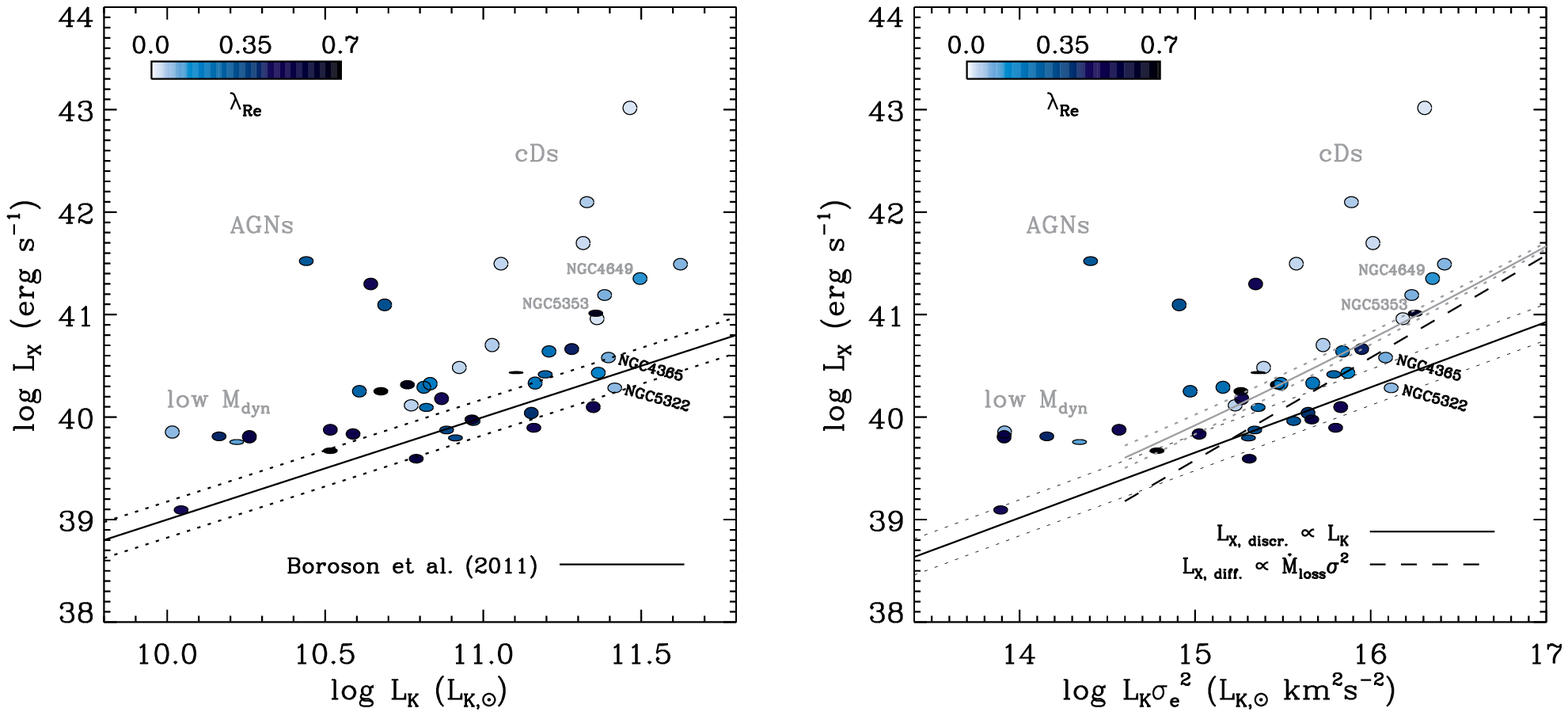}
\end{center}
\caption[]{Same as Figs.~\ref{fig:LKLX} and \ref{fig:LKsigLX}, but now
  with elliptical symbols coded by colour and flattening according to
  the degree of rotational support and the apparent ellipticity of the
  galaxies they are representing, in order to visually convey the
  dependence of the scatter in their total X-ray luminosity on these
  two physical parameters, as quantified in
  Fig.~\ref{fig:ResidCorrLX}. The grey labels indicate the galaxies
  that were excluded from the analysis of the correlation presented in
  Fig.~\ref{fig:ResidCorrLX}. The other galaxies labelled on the left
  and right panels are, respectively, most likely face-on but
  intrinsically flat and rotationally-supported galaxies and the
  flattest slow rotators in our low X-ray resolution ATLAS$^{\rm 3D}$
  subsample. 
  Finally, in both panels we also label in grey the two most massive
  X-ray bright fast-rotators, NGC~4649 and NGC~5353. The kinematic
  classification of NGC~4649 is somehow uncertain, however, since this
  object lies almost exactly on the dividing line between fast and
  slow rotators defined in Paper~III, whereas the very flat NGC~5353
  happens to be the most massive member of the compact group of
  galaxies HCG~68, so that its total $L_X$ may also include emission
  from hot gas confined by the group potential.
}
\label{fig:LKLXLKsigLXcol}
\end{figure*}
}
\newcommand{\placefigLKsigLXgas}{
\begin{figure}
\begin{center}
\includegraphics[width=\columnwidth]{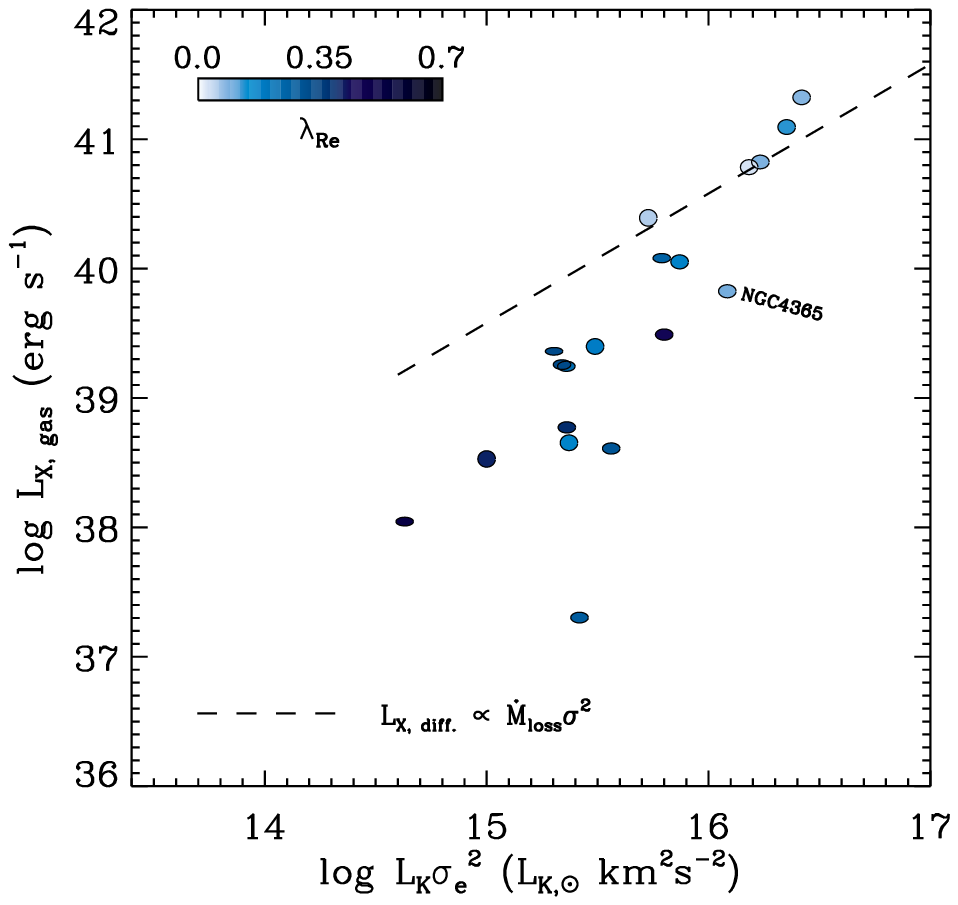}
\end{center}
\caption[]{$L_K\sigma_{\rm e}^2$ vs. $L_{X,{\rm gas}}$ diagram for our
  high X-ray resolution ATLAS$^{\rm 3D}$ sub-sample galaxies. By
  showing the X-ray luminosity $L_{X,{\rm gas}}$ due only to the hot
  gas, this diagram checks more directly than Fig.~\ref{fig:LKsigLX}
  whether the X-ray halos of the objects in our high X-ray sample
  originate from stellar ejecta that through shocks and collisions
  were heated up at X-ray emitting temperatures. The use of the same
  colour coding and squashing of Fig.~\ref{fig:LKLXLKsigLXcol} for our
  data points makes it easy to appreciate that, except for one, all
  the slow rotators in this ATLAS$^{\rm 3D}$ sub-sample agree
  remarkably well with the same simple prediction for $L_{X,{\rm
      diff}} = L_{\sigma}$ that was already shown in
  Fig.~\ref{fig:LKsigLX} and the right panel of
  Fig.~\ref{fig:LKLXLKsigLXcol}. Fast rotators, on the other hand,
  display $L_{X,{\rm gas}}$ values that fall systematically short of
  the predicted values of $L_{X,{\rm diff}}$, as does also NGC~4365,
  which is the flattest slow rotator in the high X-ray resolution
  sample.}
\label{fig:LKsigLXgas}
\end{figure}
}
\newcommand{\placefigResidCorrLXgas}{
\begin{figure*}
\begin{center}
\includegraphics[height=0.95\textwidth,angle=90]{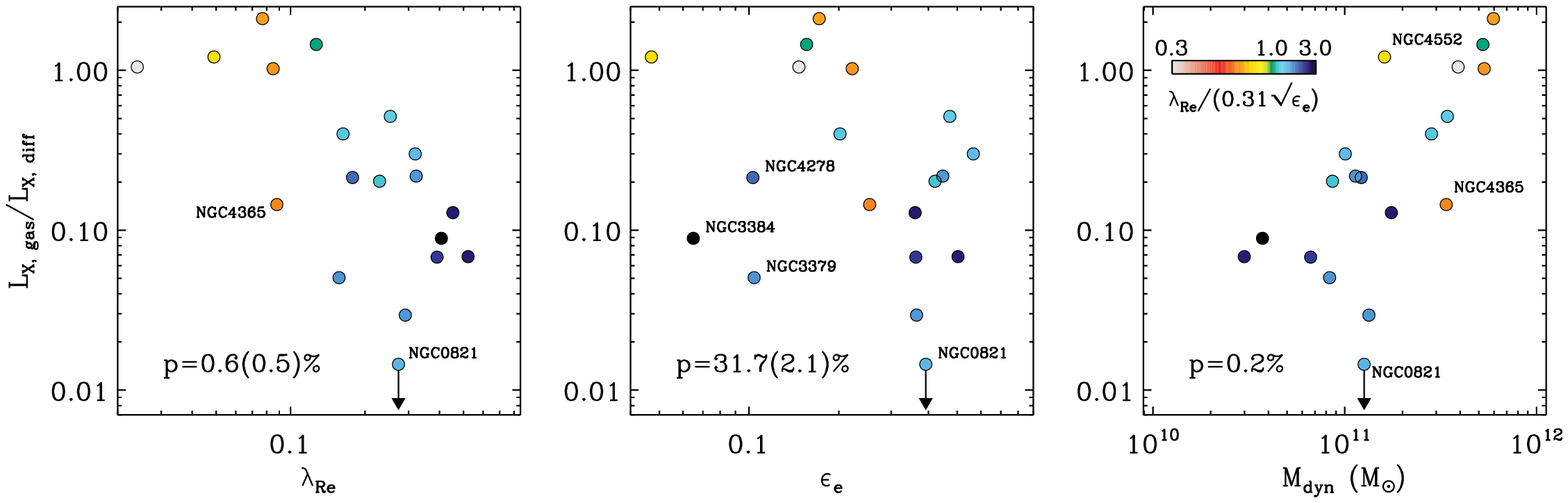}
\end{center}
\caption[]{Correlations between the observed specific angular momentum
  (quantified by the $\lambda_{\rm Re}$ parameter, left), the apparent
  flattening $\epsilon_{\rm e}$ (middle) and the dynamical mass
  $M_{\rm dyn}$ (right) of our high X-ray resolution
  \atlas\ sub-sample galaxies and their value for the ratio of the
  hot-gas luminosity $L_{X,{\rm gas}}$ to the X-ray luminosity
  $L_{X,{\rm diff}} = L_{\sigma}$ that is expected in the case of
  stellar-mass loss material heated to the kinetic temperature of the
  stars.
  As in Fig, \ref{fig:ResidCorrLX}, the significance of the
  correlation between the quantities plotted in each panel is
  indicated by the values of the Spearman null-hypothesis probability
  $p$, which were computed while excluding the exceptionally X-ray
  under-luminous galaxy NGC~821 (together with the labelled objects in
  the left and central panels in order to obtain the values of $p$
  within parentheses).
  Here the symbols are colour-coded according to their distance from
  the $\lambda_{\rm Re} = 0.31 \sqrt{\epsilon_{\rm e}}$ dividing line
  between fast and slow rotators in the $\lambda_{\rm R}$
  vs. $\epsilon$ diagram (see Fig.~7 of Paper~III), with values for
  these quantities computed within $R_{\rm e}$. Fast and slow rotators
  have $\lambda_{\rm R}$ values above and below this threshold,
  respectively.
  The $\lambda_{\rm Re}/(0.31 \sqrt{\epsilon_{\rm e}})$ ratio
  anti-correlates with the $L_{X,{\rm gas}}/L_{X,{\rm diff}}$ deficit
  as significantly as $\lambda_{\rm Re}$ or $M_{\rm dyn}$, with a
  null-hypothesis probability $p=0.3\%$ (or 0.1\% without NGC~4365).}
\label{fig:ResidCorrLXgas}
\end{figure*}
}
\newcommand{\placefigeight}{
\begin{figure*}
\begin{center}
\includegraphics[width=0.95\textwidth]{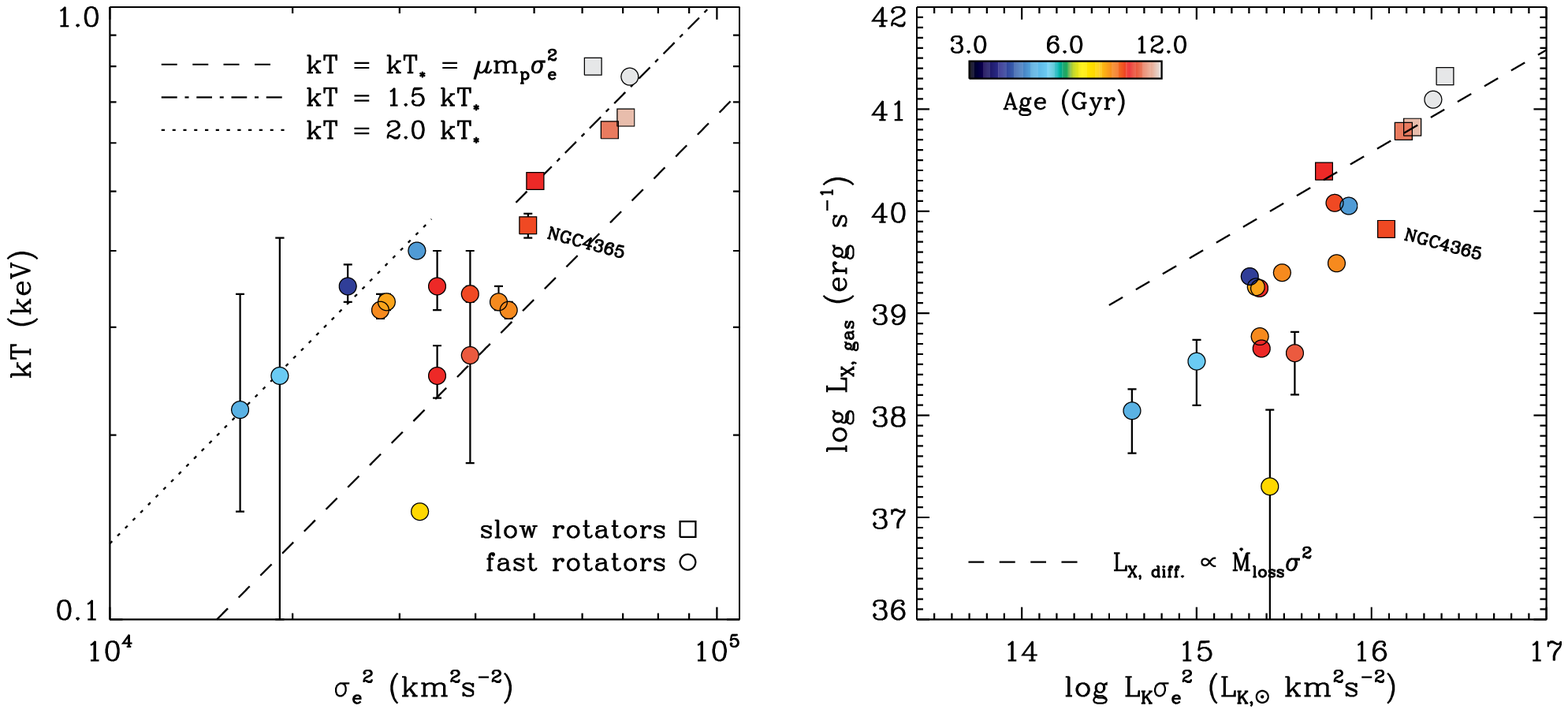}
\end{center}
\caption[]{Left: the hot-gas temperature $kT$ of the objects in our
  high X-ray resolution ATLAS$^{\rm 3D}$ sub-sample
  \citep[from][]{Bor11} against the square of their stellar velocity
  dispersion measured within one effective radius, $\sigma_{\rm
    e}$. Square and round symbols indicate slow and fast rotators,
  respectively and according to the $\lambda_{\rm Re} = 0.31
  \sqrt{\epsilon_{\rm e}}$ dividing line introduced in Paper~III,
  whereas the colour-coding traces the average stellar age of our
  sub-sample galaxies (from McDermid et al. in preparation), again
  within one $R_e$. The dashed line indicates the expected temperature
  if the stellar motions are the main heat source for the hot gas. The
  $kT$ values of slow rotators appear consistent with such a scenario,
  albeit while allowing for an extra $\sim50\%$ energy input (shown by
  the dashed-dotted line). Fast rotators, on the other hand, show
  similar $kT$ values between 0.25 and 0.35 keV across a range of
  $\sigma_{\rm e}^2$ values, although the youngest objects appear to
  show systematically larger temperature values (by a factor two,
  dotted line).
  Right: same $L_K\sigma_{\rm e}^2$ vs. $L_{X,{\rm gas}}$ diagram as
  Fig.~\ref{fig:LKsigLXgas}, but now using the same symbols and
  colour-coding as in the left panel in order to illustrate how the
  youngest fast rotators may also have brighter X-ray haloes.
  The position of the flat slow rotator NGC~4365 is indicated in both
  panels to show that despite being X-ray under-luminous, this galaxy
  appears to display a hot-gas temperature well in line with the rest
  of the objects in its kinematic class.
  The $kT$ estimate for NGC~821 (yellow point) is very uncertain, and
  errors are not shown for clarity. The most massive fast rotator in
  both panels is again NGC~4649, whose uncertain kinematic
  classification was noticed in Fig.~\ref{fig:LKLXLKsigLXcol}.  }
\label{fig:kTandAges}
\end{figure*}
}
\newcommand{\placefignine}{
\begin{figure}
\begin{center}
\includegraphics[width=\columnwidth]{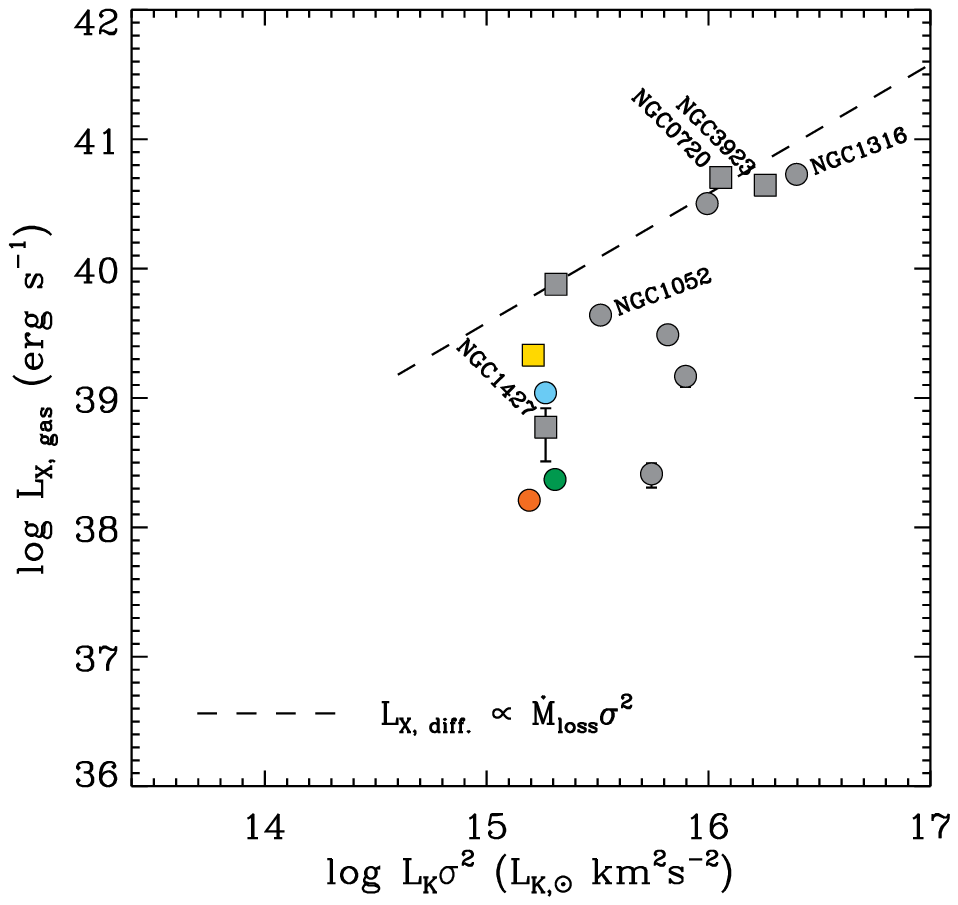}
\end{center}
\caption[]{$L_K\sigma^2$ vs. $L_{X,{\rm gas}}$ diagram, similar to
  Fig.~\ref{fig:LKsigLXgas} and the right-hand panel of
  Fig.~\ref{fig:kTandAges}, but now showing either the remaining
  objects from \citet[][except M~32]{Bor11} that are not in the
  \atlas\ sample (grey points) or galaxies with \Chandra\ measurements
  of $L_{X,{\rm gas}}$ and \atlas\ data \citep[coloured points,
    from][]{Mul10,Ath07}. The Boroson et al. objects are plotted using
  their values for $L_{X,{\rm gas}}$, $L_K$ and the central value for
  the stellar velocity dispersion $\sigma$, whereas for the additional
  \atlas\ galaxies we used our $L_K$ and $\sigma_{\rm e}$ measurements
  and rescaled $L_{X,{\rm gas}}$ using our adopted distance
  estimates. This also applies to NGC~720, for which \sauron\ data
  exist. Similarly to Fig.~\ref{fig:kTandAges}, square and round
  symbols indicate slow and fast rotators, respectively (see text for
  details on how such a separation was done for the Boroson et
  al. galaxies), with objects in the \atlas\ sample being colour-coded
  according to our estimates for the average stellar age (from
  McDermid et al. in preparation). Finally, the labels indicate either
  objects with clear signs of recent star formation or particularly
  flat slow rotators (when inclined to the right and left,
  respectively). These additional data further suggest that the
  discrepancy in the hot-gas content of fast- and slow rotators may
  reduce or even disappear in very massive galaxies and that the
  presence of younger stellar sub-populations may increase the hot-gas
  luminosity, while also revealing that galactic flattening may be a
  necessary but not sufficient reason for the $L_{X,{\rm gas}}$
  deficiency observed in a minority of slow rotators.}
\label{fig:moreLKsigLXgas}
\end{figure}
}
\newcommand{\placefigten}{
\begin{figure}
\begin{center}
\includegraphics[width=\columnwidth]{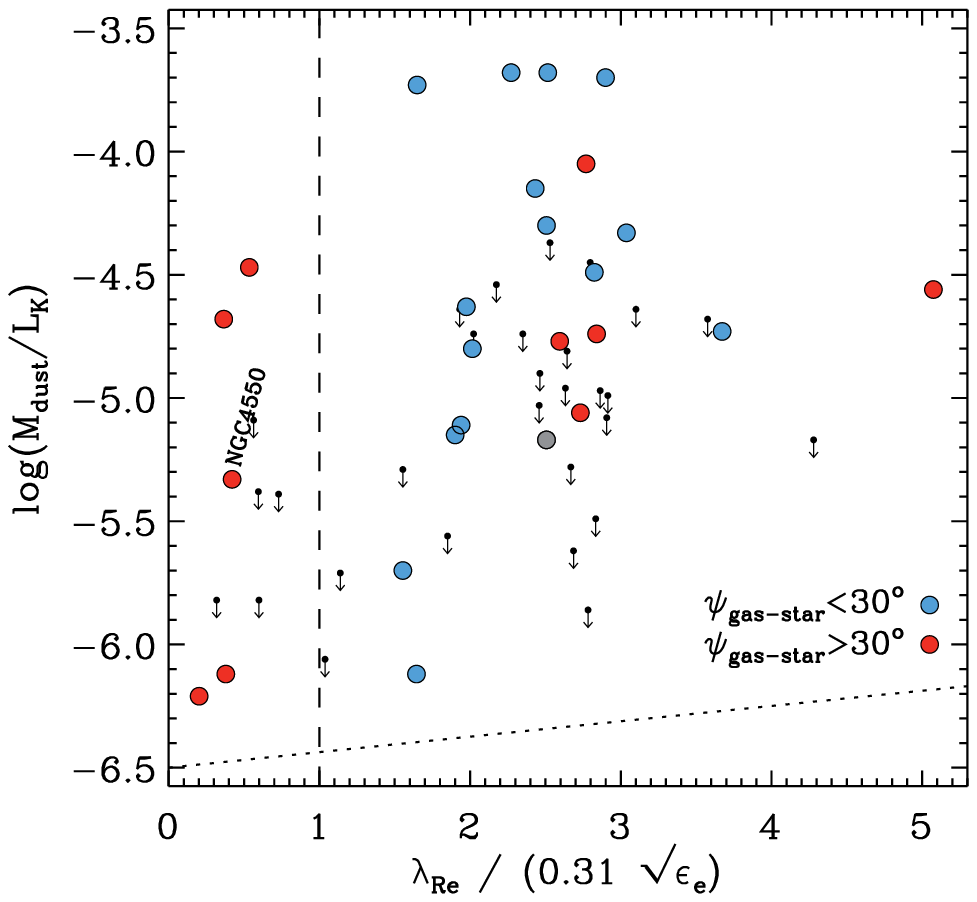}
\end{center}
\caption[]{Specific dust mass content $\log{(M_{\rm dust}/L_K)}$ for
  the \atlas\ sample galaxies with {\it Herschel\/} data analysed by
  \citet{Smi12} against their degree of rotational support, as
  quantified by the value of the ${\rm \lambda_{\rm
      Re}/(0.31\sqrt{\epsilon_{\rm e}}})$ ratio that is also used to
  separate slow from fast rotators (vertical dashed line). The filled
  circles denote galaxies with $M_{\rm dust}$ detections, whereas the
  downward pointing arrows indicate objects with just $M_{\rm dust}$
  upper-limits. The dotted line shows the detection limit on the
  specific dust mass for this sub-sample, reflecting the fact that
  limits as low as $\log{M_{\rm dust}}=5.0$ can be set across the
  whole considered range of ${\rm \lambda_{\rm
      Re}/(0.31\sqrt{\epsilon_{\rm e}}})$ values whereas fast rotators
  extend to lower maximum $\log{L_K}$ luminosities than in the case of
  slow rotators. Except in one case (shown in grey), the objects with
  detected $M_{\rm dust}$ values also display diffused ionised-gas
  emission, for which \citet{Dav11} measured the angle between the
  angular momentum of the stars and the ionised-gas. Objects with
  aligned ionised-gas and stellar motions are shown in blue, whereas
  objects with a kinematic misalignment greater than $30^{\circ}$ are
  shown in red. Such a kinematic information suggests that the
  ionised-gas and dust of slow rotators is accreted, whereas most of
  the dustier fast rotators (i.e., with $M_{\rm dust}/L_K$ values in
  excess of what can be found in slow rotators) show aligned stellar
  and ionised-gas motions, consistent with an internal origin of both
  the warm gas and dust. As ${\rm \lambda_{\rm
      Re}/(0.31\sqrt{\epsilon_{\rm e}}})$ traces well the $L_{X,{\rm
      gas}}$ deficiency of fast rotators (\S~3.2), their ability to
  preserve more of their dusty stellar mass-loss material could relate
  to the presence of a more tenuous hot-gas medium compared to the
  case of slow rotators. In this respect, we note that the special
  slow-rotator NGC~4550 \citep{Ems07} is very likely to be $L_{X,{\rm
      gas}}$ deficient given that it is intrinsically flat.}
\label{fig:Herschel}
\end{figure}
}

\title[The X-ray Halos of Fast and Slow Rotators]{The ATLAS$^{\rm3D}$
  project - XIX. The hot-gas content of early-type galaxies: fast
  versus slow rotators}

\author[Sarzi et al.]{Marc Sarzi$^{1}$\thanks{E-mail :sarzi@star.herts.ac.uk}\\
  $^{1}$Centre for Astrophysics Research, University of Hertfordshire, 
  College Lane, Hatfield, Herts, AL10 9AB, UK\\}

\author[Sarzi et al.] {\parbox{\textwidth}{Marc
    Sarzi,$^{1}$\thanks{E-mail:\texttt{ m.sarzi@herts.ac.uk}}
Katherine Alatalo,$^{2}$ Leo Blitz,$^{2}$ Maxime Bois,$^{3}$
Fr\'ed\'eric Bournaud,$^{4}$ Martin Bureau,$^{5}$ Michele
Cappellari,$^{5}$ Alison Crocker,$^{6}$ Roger L. Davies,$^{5}$ Timothy
A. Davis,$^{7}$ P. T. de Zeeuw,$^{7,8}$ Pierre-Alain Duc,$^{4}$ Eric
Emsellem,$^{7,9}$ Sadegh Khochfar,$^{10}$ Davor Krajnovi\'c,$^{7}$
Harald Kuntschner,$^{11}$ Pierre-Yves Lablanche,$^{2,9}$ Richard
M. McDermid,$^{12}$ Raffaella Morganti,$^{13,14}$ Thorsten
Naab,$^{15}$ Tom Oosterloo,$^{13,14}$ Nicholas Scott,$^{16}$ Paolo
Serra,$^{13}$ Lisa M. Young,$^{17}$\thanks{Adjunct Astronomer with
  NRAO} and Anne-Marie Weijmans,$^{18}$\thanks{Dunlap
  Fellow}}\vspace{0.4cm}\\
\parbox{\textwidth}{
$^{1}$Centre for Astrophysics Research, University of Hertfordshire,
  Hatfield, Herts AL1 9AB, UK\\
$^{2}$Department of Astronomy, Campbell Hall, University of
  California, Berkeley, CA 94720, USA\\
$^{3}$Observatoire de Paris, LERMA and CNRS, 61 Av. de l`Observatoire,
  F-75014 Paris, France\\
$^{4}$Laboratoire AIM Paris-Saclay, CEA/IRFU/SAp -- CNRS --
  Universit\'e Paris Diderot, 91191 Gif-sur-Yvette Cedex, France\\
$^{5}$Sub-Department of Astrophysics, Department of Physics,
  University of Oxford, Denys Wilkinson Building, Keble Road, Oxford,
  OX1 3RH\\
$^{6}$Department of Astronomy, University of Massachussets, Amherst, MA 01003, USA\\
$^{7}$European Southern Observatory, Karl-Schwarzschild-Str. 2, 85748
  Garching, Germany\\
$^{8}$Sterrewacht Leiden, Leiden University, Postbus 9513, 2300 RA
  Leiden, the Netherlands\\
$^{9}$Universit\'e Lyon 1, Observatoire de Lyon, Centre de Recherche
  Astrophysique de Lyon and Ecole Normale Sup\'erieure de Lyon, 9
  avenue Charles Andr\'e, F-69230 Saint-Genis Laval, France\\
$^{10}$Max-Planck Institut f\"ur extraterrestrische Physik, PO Box
  1312, D-85478 Garching, Germany\\
$^{11}$Space Telescope European Coordinating Facility, European
  Southern Observatory, Karl-Schwarzschild-Str. 2, 85748 Garching,
  Germany\\
$^{12}$Gemini Observatory, Northern Operations Centre, 670 N. A`ohoku
  Place, Hilo, HI 96720, USA\\
$^{13}$Netherlands Institute for Radio Astronomy (ASTRON), Postbus 2,
  7990 AA Dwingeloo, The Netherlands\\
$^{14}$Kapteyn Astronomical Institute, University of Groningen,
  Postbus 800, 9700 AV Groningen, The Netherlands\\
$^{15}$Max-Planck-Institut f\"ur Astrophysik,
  Karl-Schwarzschild-Str. 1, 85741 Garching, Germany\\
$^{16}$Centre for Astrophysics \& Supercomputing, Swinburne University
  of Technology, PO Box 218, Hawthorn, VIC 3122, Australia\\
$^{17}$Physics Department, New Mexico Institute of Mining and
  Technology, Socorro, NM 87801, USA\\
$^{18}$Dunlap Institute for Astronomy \& Astrophysics, University of
  Toronto, 50 St. George Street, Toronto, ON M5S 3H4, Canada
}
}

\begin{document}
\pagerange{\pageref{firstpage}--\pageref{lastpage}} \pubyear{2012}

\maketitle
\label{firstpage}

\clearpage

%
\begin{abstract}

For early-type galaxies, the ability to sustain a corona of hot, X-ray
emitting gas could have played a key role in quenching their
star-formation history. A halo of hot gas may act as an effective
shield against the acquisition of cold gas and can quickly absorb
stellar-mass loss material.  Yet, since the discovery by the
\Einstein\ observatory of such X-ray halos around early-type galaxies,
the precise amount of hot gas around these galaxies still remains a
matter of debate.
By combining homogeneously-derived photometric and spectroscopic
measurements for the early-type galaxies observed as part of the
\atlas\ integral-field survey with measurements of their X-ray
luminosity based on X-ray data of both low and high spatial resolution
(for 47 and 19 objects, respectively) we conclude that the hot-gas
content of early-type galaxies can depend on their dynamical
structure.
Specifically, whereas slow rotators generally have X-ray halos with
luminosity $L_{X,{\rm gas}}$ and temperature $T$ values that are well
in line with what is expected if the hot-gas emission is sustained by
the thermalisaton of the kinetic energy carried by the stellar-mass
loss material, fast rotators tend to display $L_{X,{\rm gas}}$ values
that fall consistently below the prediction of this model, with
similar $T$ values that do not scale with the stellar kinetic energy
(traced by the stellar velocity dispersion) as observed in the case of
slow rotators.
Such a discrepancy between the hot-gas content of slow and fast
rotators would appear to reduce, or even disappear, for large values
of the dynamical mass (above $\sim3\times10^{11}M_{\odot}$), with
younger fast rotators displaying also somewhat larger $L_{X,{\rm
    gas}}$ values possibly owing to the additional energy input from
recent supernovae explosions.
Considering that fast rotators are likely to be intrinsically flatter
than slow rotators, and that the few $L_{X,{\rm gas}}$-deficient slow
rotators also happen to be relatively flat, the observed $L_{X,{\rm
    gas}}$ deficiency in these objects would support the hypothesis
whereby flatter galaxies have a harder time in retaining their hot
gas, although we suggest that the degree of rotational support could
further hamper the efficiency with which the kinetic energy of the
stellar-mass loss material is thermalised in the hot gas.
We discuss the implications that a different hot-gas content could
have on the fate of both acquired and internally-produced gaseous
material, considering in particular how the $L_{X,{\rm gas}}$
deficiency of fast rotators would make them more capable to recycle
the stellar-mass loss material into new stars than slow rotators. This
would be consistent with the finding that molecular gas and young
stellar populations are detected only in fast rotators across the
entire \atlas\ sample, and that fast rotators tend to have a larger
specific dust mass content than slow rotators.

\end{abstract}

%

%
\section{Introduction}
\label{sec:intro}

Early-type galaxies used to be regarded as purely stellar systems over
the largest part of the last century, even though the presence of
little or no interstellar medium in these old stellar systems was soon
recognised as a problem by \citet{Fab76}. Indeed, such a lack of gas
contrasts with the notion that old stars lose a considerable fraction
of their initial mass during their evolution.
In this context, the finding by the \Einstein\ observatory that
early-type galaxies are surrounded by massive haloes of hot ($\sim
10^6-10^7$K) X-ray emitting gas \citep[e.g.][]{For79} both changed our
view of early-type galaxies and provided a possible solution to the
problem posed by \citeauthor{Fab76}, if a mechanism to heat the
stellar ejecta to X-ray temperatures could be found.
Around the same time, the advent of new and homogeneous sets of
optical data for large numbers of early-type galaxies prompted the
finding of a correlation between the X-ray luminosity of such haloes
$L_X$ and the optical luminosity $L_B$ of the galaxies they contain
\citep[e.g.][]{For85,Tri85}, which when interpreted led to a
relatively simple model linking the hot gas reservoirs to the
stellar-mass loss material of early-type galaxies
\citep[e.g.][]{Can87}.
According to this scenario, still part of the current consensus, the
gas ejected from evolved stars and planetary nebulae collides and
shocks with other ejecta or a pre-existing ambient gas until it is
heated to the kinetic temperature that corresponds to the stellar
velocity dispersion of a galaxy.
Supernovae explosions \citep[in particular of type Ia,
  e.g.][]{Mat86,Cio91} contribute to further heat this gas, but may
lead also to its escape.
This would have generally been the case when galaxies were young and
SNe explosions more frequent, whereas today SNe-driven global outflows
should be more restricted to the less massive systems.
On the other hand, bigger galaxies can hold better to their hot gas,
and in amounts that scale with their optical luminosity given the
stellar origin of such a material.

Yet, despite its tremendous success in explaining the broad trend of
the $L_B-L_X$ relation, this idea alone does not account for the wide
range of X-ray luminosities, of about two orders of magnitude, that is
observed in galaxies of similar optical luminosity.
A number of physical phenomena have been examined to account for the
scatter of the $L_B-L_X$ relation (see, e.g., the reviews of
\citealp{Mat03} and \citealp{Pel12}).
Earlier explanations invoked different degrees of ram-pressure
stripping, since galaxies in dense galactic environment were found to
be X-ray faint \citep{Whi91}, or different stages of the hot-gas
evolution caused by the steady decrease over time of the energy
provided by SNe or the thermalisation of the stellar motions, which by
eventually falling short of the power required to extract the gas from
a galaxy drives galaxies from an X-ray faint wind and outflow phase to
an X-ray bright inflow phase \citep[and at different speed for
  galaxies with different structural parameters such as central
  density, effective radius or kind of radial profile;][]{Cio91}.
Subsequently, the finding by \citet{Esk95a,Esk95b} that S0s and flat
Es show lower X-ray luminosities than rounder elliptical galaxies of
the same optical luminosity prompted further theoretical studies
concerning the role of intrinsic flattening \citep[which reduces the
  binding energy of the hot gas and makes it harder to retain
  it;][]{Cio96} or rotation \citep{DEr98,Bri96}.
More recent studies considered also the impact of secondary gas
accretion events \citep{Pip05} and showed the importance of
contamination from the radiation of the intracluster medium
\citep[ICM, which originates most likely from cosmological
  flows;][]{Mat03} in the $L_X$ measurements obtained from relatively
shallow surveys \citep{Mat01}.

The launch of the \Chandra\ and \XMM\ space telescopes, with their
higher spatial resolution, spectral range and sensitivity, has now
made it possible to isolate the X-ray emission of galactic haloes from
the signal of unresolved low-mass X-ray binaries (LMXBs), active
nuclei (AGNs) or the ICM that otherwise enters the coarser
\Einstein\ or \ROSAT\ measurements of $L_X$.
These measurements are very expensive in terms of telescope time,
however, and as a result the number of objects with pure $L_{X,{\rm
    gas}}$ values for hot-gas haloes is still quite limited. This is
particular true considering that good quality and homogeneous optical
data, as well as robust distance estimates, are also highly desirable
in order to understand what drives the scatter in the X-ray properties
of galaxies.
On the other hand, since also photometric and spectroscopic optical
measurements for nearby galaxies are witnessing a quantum leap both in
quality and sky coverage, it may be possible to make progress in this
domain while still using $L_X$ measurements based on \ROSAT\ or
\Einstein\ data, which cover a large number of objects.
This was for instance the case of the study of \citet{Ell06}, who used
the near-infrared measurements from the Two-Micron All Sky Survey
\citep[2MASS;][]{Skr06} to revisit the stellar to X-ray luminosity
relation while still using nearly all of the 401 early-type galaxies
in the catalog of \citet{oSu01}, still as of today the largest
compilation of $L_X$ measurements obtained with \ROSAT\ and \Einstein.

By drawing on new optical data from the complete \atlas\ survey
\citep[][hereafter Paper~I]{Cap11a}, in this paper we will use both
such approaches to reconsider the origin of the scatter in the X-ray
luminosity of early-type galaxies, starting from \ROSAT\ or
\Einstein\ data for the total X-ray luminosity of 47 galaxies from the
\atlas\ survey to then analyse a second \atlas\ sub-sample of 19
objects with consistent \Chandra\ measurements of the X-ray luminosity
of their hot gas.
In particular, we will revisit the possible role of galactic rotation
and flattening in driving the hot-gas content of galaxies, building on
the first observational tests of \citet{Pel97} and in light of the
emerging distinction among early-type galaxies between a dominant
population of fast-rotating objects \citep[with regular velocity
  fields;][hereafter Paper~II]{Kra11} and a less common class of
slowly-rotating galaxies \citep[often hosting kinematically distinct
  cores;][hereafter Paper~III]{Ems11}.
Fast and slow rotators are indeed not only likely to have formed in
different ways \citep[][]{Cap07,Ems07} but also to have distinct
distributions for their intrinsic shapes, with fast rotators being
nearly as flat as spiral galaxies and slow rotators being much rounder
and possibly triaxial \citep[Weijmans et al. in preparation, but see
  already][hereafter Paper~VII]{Cap11b}.

This paper is organised as follows. 
In \S~\ref{sec:data} we define our two samples and the optical
parameters from the \atlas\ survey that will be needed for our
investigation.
We derive our results in \S~\ref{sec:results}, considering first in
\S~\ref{subsec:resultsLow} an \atlas\ sub-sample with X-ray
measurements based on \ROSAT\ and \Einstein\ data.
We then expand our analysis in \S~\ref{subsec:resultsHigh} to
\atlas\ galaxies with \Chandra\ data, and further check in
\S~\ref{subsec:resultsMore} the results from both
\S~\ref{subsec:resultsLow} and \S~\ref{subsec:resultsHigh} using an
additional but less homogeneous set of \Chandra\ X-ray measurements
from the literature.
We discuss our findings in \S~\ref{sec:discussion}, and wrap up our
conclusions in \S~\ref{sec:conc}.

\section{X-ray Data and Samples Properties}
\label{sec:data}

The two sets of galaxies that will be used in this work simply consist of: 
a) all the objects in the \atlas\ sample in Paper~I for which
\citet{oSu01} provides a \ROSAT\ or \Einstein\ measurement for their
$L_X$ (considering only detections), and
b) of all the \atlas\ sample galaxies for which \citet[][hereafter
  also Boroson et al.]{Bor11} could use \Chandra\ data to separate the
$L_X$ contribution from the otherwise unresolved population of LMXBs
and from the hot, X-ray emitting gas.
These two \atlas\ sub-samples include 47 and 19 galaxies,
respectively, and since the main difference between them is the
spatial resolution of the X-ray data from the literature that were
used to compile them, hereafter we will refer to these two sets of
objects as the low and high X-ray resolution samples.
Furthermore, we note that whereas O'Sullivan et al. include all sort
of early-type galaxies, Boroson et al. restricted themselves to well
studied X-ray objects and explicitly excluded the dominant galaxies of
groups and clusters that are associated to extended hot gas haloes
confined by the group or cluster gravitational potential.

For the galaxies in these two \atlas\ sub-samples Paper~I provides
their currently best distance estimates, consistently derived values
for their K-band luminosity $L_K$ starting from 2MASS apparent
magnitude, and effective radius $R_{\rm e}$.
Homogeneously derived values for the apparent ellipticity $\epsilon =
1-b/a$ were obtained in Paper~III through the analysis of the images
at our disposal (mostly from the Sloan Digital Sky Survey), whereas
\citet[][hereafter Paper~XV]{Cap12b} provide us with accurate
dynamical mass measurement, $M_{\rm dyn}$, based on the detailed Jeans
anisotropic dynamical modelling \citep{Cap08} of the stellar
kinematics obtained from our \sauron\ data.\footnote{
  More specifically, the values used here are those based on the
  self-consistent model (B) of Paper~XV, where they are indicated as
  $M_{\rm JAM}$. Such $M_{\rm JAM}$ values provide an approximation
  for twice the mass contained within a sphere enclosing half of the
  galaxy light. Given that the stars dominate the mass budget inside
  that sphere, $M_{\rm JAM}$ also provides a good estimate for the
  total stellar mass $M_{\rm stars}$ of a galaxy, while also including
  possible variations of the initial stellar mass function \citep[see,
    e.g.,][]{Cap12a}.}

The {\tt SAURON\/} data also allow us to probe the depth of the
gravitational potential well using values for the stellar velocity
dispersion $\sigma$ extracted within the same physical aperture and to
assess the degree of rotational support using the $\lambda_{\rm R}$
parameter introduced by \citet{Ems07}.
Since we are interested in the global shape and degree of rotational
support of our sample galaxies, we adopt here the values for
$\lambda_{\rm Re}$ and $\epsilon_{\rm e}$ that in Paper~III were
computed within the galaxy effective radius $R_{\rm e}$.
Similarly, as we are interested in tracing most of the potential well,
we will use the stellar velocity dispersion values $\sigma_{\rm e}$
(also provided by Paper~XV) that are extracted within an elliptical
aperture of equivalent circular radius $R_{\rm e}$.
The use of $\epsilon$ in conjunction with $\lambda_{\rm R}$ will help
us in separating face-on but intrinsically flat and fast-rotating
objects from much rounder and slowly-rotating galaxies, as well as in
identifying galaxies that may be intrinsically quite flat also in this
latter class of objects (most likely as they are triaxial in shape).

All the quantites described in this section are listed in Tabs.~1 and 2
for the galaxies in the low and high X-ray resolution subsamples,
respectively. These tables also include rescaled X-ray luminosity
values from the literature assuming the \atlas\ distance estimates,
which, unless otherwised stated, will be used in our analysis. {\it
  Note: These tables will appear on the present arXiv version of the
  paper as soon as Paper~XV will also be posted on it.}

%
\section{Results}
\label{sec:results}

\subsection{Low X-ray Resolution Sample}
\label{subsec:resultsLow}

\placefigLBLX

We start this section by showing in Fig.~\ref{fig:LBLX} how the
galaxies in the low X-ray resolution sample fare in the classic
$L_B-L_X$ diagram when using the values for the B-band and X-ray
luminosity taken from the catalogue of \citet{oSu01}, without any
distance rescaling.
The dashed and solid lines in Fig.~\ref{fig:LBLX} trace the expected
contribution to the observed $L_X$ from the unresolved emission of
low-mass X-ray binaries, as first estimated by O'Sullivan et al. and
subsequently by \citet{Kim04}.
The dotted lines also show the uncertainties associated with the
\citeauthor{Kim04} calibration, which was based on the luminosity
function of LMXBs resolved with \Chandra\ images in a sample of 14 E
and S0s.
The other early-type galaxies with $L_X$ detections in the O'Sullivan
et al. catalogue are also plotted in Fig.~\ref{fig:LBLX}, to
illustrate better how traditionally a
$L_B\sim1-3\times10^{10}L_{\odot}$ would mark the onset of a $L_X
\propto L_B^2$ trend due to the presence of a diffuse component.
X-ray bright objects at low $L_B$ values in this diagram are most
likely outliers that own their X-ray flux either to a central AGN or
the surrounding ICM, which should be isolated when considering the
nature of the scatter in the $L_X$ values from the hot gas.
In fact, the three labelled sources in Fig.~\ref{fig:LBLX} were
already identified by O'Sullivan et al. as powerful AGNs.
Fig.~\ref{fig:LBLX} also shows that due to the volume-limited nature
of the \atlas\ sample our objects fall short from covering the entire
range of optical luminosity spanned by the sample of O'Sullivan et al.,
although only by a factor two.

\placefigLKLX

We now move on from this first diagram and start making use of our
\atlas\ database by considering in Fig.~\ref{fig:LKLX} the
near-infrared K-band luminosity of our sample galaxies, which
corresponds much more closely to the total stellar mass than does the
B-band optical luminosity, in particular in the occasional presence in
early-type galaxies of a younger stellar sub-component.
Furthermore, in Fig.~\ref{fig:LKLX} we now adopt the currently best
estimates for the distance of our sample galaxies, not only to compute
the $L_K$ luminosity from the 2MASS apparent K-band magnitudes but
also to rescale the $L_X$ values given in O'Sullivan et al..
Compared to Fig.~\ref{fig:LBLX}, in the $L_K-L_X$ diagram the bright
galaxies with the largest X-ray luminosities appear to distance
themselves more from other objects with $L_X$ exceeding the value
expected from discrete sources.
All such galaxies are either deeply embedded in the intracluster
medium of the Virgo cluster (NGC~4486, NGC4406) or are the central
member of their own group of galaxies and show a rather extended X-ray
halo \citep[NGC~4636 and NGC~5846;][]{Mat01,Mul03,San07}.
This suggests that the ICM contributes largely to the $L_X$ values of
these objects, which should then be flagged in the remainder of our
analysis as in the case of the AGNs identified above.
We also note that in the $L_K-L_X$ diagram $L_K$ is a more accurate
predictor for the contribution of LMXBs to the total $L_X$ values,
although whether most or just a few of our sample objects could be
considered to possess an X-ray halo still depends, like in
Fig.~\ref{fig:LBLX}, on the adopted calibration for the $L_X$ from
LMXBs.
Yet, the \citet{Kim04} calibration appears problematic since according
to it there would appear to be a considerable fraction of galaxies in
the $L_K-L_X$ diagram whose $L_X$ values are nearly half what would be
expected from LMXBs alone.
The more recent calibration of \citet{Bor11} does not suffer from this
bias, which is why we will adopt it throughout the rest of our
analysis.

\placefigLKsigLX

To further understand the nature of the $L_X$ emission in our sample
galaxies, in Fig.~\ref{fig:LKsigLX} we have placed them in the
$L_K\sigma_{\rm e}^2$ vs. $L_X$ diagram.
This is a near-infrared and integral-field version of the
$L_B\sigma^2$ vs. $L_X$ diagram introduced by Canizares et al. (1987),
where the combination of $L_K$ and $\sigma_{\rm e}$ allows to trace
the X-ray luminosity that is expected if the energy lost by the hot
gas through radiation is compensated by the thermalisation of
the kinetic energy that the stellar ejecta inherit from their parent
stars.
Such an input rate of kinetic energy $L_{\sigma}$ would indeed
correspond to the stellar-mass loss rate $\dot M$ (traced by $L_K$)
times the specific kinetic energy associated with the stellar random
motions (traced by $\sigma_{\rm e}^2$). More specifically, we take
$L_{\sigma} = \frac{3}{2}\dot M \sigma_{\rm e}^2$.
To show the prediction of such a model in Fig.~\ref{fig:LKsigLX} we
computed the average ratio between the K- and B-band luminosity of our
sample galaxies ($\log{L_K/L_B}=0.7$) and assumed a mass-loss rate
$\dot M$ of $1.5 \times 10^{-11} (L_B/L_{\odot}) t_{15}^{-1.3}
M_{\odot} \rm yr^{-1}$ \citep[as given by][]{Cio91} with an average
stellar age $t_{15}$ (in units of 15 Gyr) of 12 Gyr.
By adding to this model the expected contribution $L_{X,{\rm discr}}$
from the unresolved LMXB population according to the calibration of
Boroson et al., which we plotted in the $L_K\sigma_{\rm e}^2-L_X$
diagram using the best fitting near-infrared Faber-Jackson correlation
\citep{FJ76} between the $L_K$ and $\sigma_{\rm e}$ values for our low
X-ray resolution sample ($\log{\sigma_{\rm e}} =
0.285\log{L_K}-0.864$), we finally obtain the total $L_X$ values that
ought to be compared with the data shown in Fig.~\ref{fig:LKsigLX}.

Incidentally, except for two objects, all the slowly-rotating galaxies
in our low X-ray resolution sample that are massive enough to heat up
the stellar-loss material at X-ray temperatures and lead to detectable
X-ray luminosities against the LMXBs background (in
Fig.~\ref{fig:LKsigLX} we considered a generous lower limit for
$\log{L_K\sigma_{\rm e}^2}$ of $14.6 \,\rm L_{K,\odot}\,km^2\,s^{-2}$,
corresponding to $M_{\rm dyn} \ga 2\times10^{10}\,\rm M_{\odot}$)
align themselves along the lines of such a simple model, falling
either directly on top or just above the predicted $L_X$ values.
We note that a situation where the radiation losses of the hot
atmosphere are balanced by the thermalisation of the kinetic energy of
the stellar ejecta would implicitely require that the stellar-mass
loss material is steadily removed from the galaxy, in order the
prevent the accumulation of gas and catastrophic cooling.
For a typical old and massive early-type galaxy such a quasi-static
situation could be maintained by SNe type Ia explosions, since at
their present-day rate the SNe energy output $L_{\rm SN}$ is
comparable to the energy $L^{-}_{\rm grav}$ that is necessary to
extract the stellar-mass material \citep[see, e.g., the spherical
  testcase of][with a $L_B = 5\times 10^{10}L_B$ that would correspond
  to $\log{L_K\sigma_{\rm e}^2} \sim 16.2$]{Pel12}.

On the other hand, most of the fast-rotating galaxies in this sample
fall short of the X-ray luminosity expected from stellar ejecta that
through shocks and collisions contribute to heat the hot gas.
If we consider the early finding of \citet{Esk95a,Esk95b} that S0s and
flat Es have a lower X-ray luminosity compared to rounder elliptical
of similar optical luminosity, and that when early-type galaxies are
split by their degree of rotational support 94\% of S0s and 66\% of Es
fall in the fast-rotating category (Paper~III), then it seems
natural to ask whether the apparent X-ray deficiency observed in
Fig.~\ref{fig:LKsigLX} for this class of objects is actually driven by
the $\lambda_{\rm R}$ parameter or by the apparent flattening
$\epsilon$.

\placefigResidCorrLX
\placefigLKLXandLKsigLXcol

For the objects in our low X-ray resolution \atlas\ sub-sample,
Fig.~\ref{fig:ResidCorrLX} shows how the ratio of their total X-ray
luminosity $L_X$ to the expected contribution from unresolved discrete
sources only $L_{X,{\rm discr}}$ or to the predicted X-ray luminosity
when including also the emission from stellar-mass loss material
heated to the stellar kinetic temperature $L_{X,{\rm discr+diff}}$
correlates with the $\lambda_{\rm Re}$ parameter, apparent flattening
$\epsilon_{\rm e}$ and dynamical mass $M_{\rm dyn}$.
A Spearman rank analysis reveals that both the $L_X/L_{X,{\rm discr}}$
and $L_X/L_{X,{\rm discr+diff}}$ ratios correlate with a modest degree
of significance with both $\lambda_{\rm Re}$ and $\epsilon_{\rm e}$
(see Fig.~\ref{fig:ResidCorrLX} for the probability of the null
hypothesis for each correlation). 
The $L_X/L_{X,{\rm discr}}$ excess would appear to be also somehow
correlated to the dynamical mass $M_{\rm dyn}$, although no such trend
is observed in the case of the $L_X/L_{X,{\rm discr+diff}}$ deficit.
In other words, pretty much independently of their dynamical mass,
increasingly rounder and more slowly-rotating galaxies in our low
X-ray resolution \atlas\ sub-sample appear to show also larger $L_X$
values that progressively exceed what is expected from the LMXB
population and eventually become consistent with the total predicted
X-ray luminosity from LMXBs and shock-heated stellar ejecta.
Conversely, the flatter and more rotationally-supported a galaxy, the
larger its $L_X$ deficit from the expectation of this simple model.

The most notable outliers from the trends observed in
Fig.~\ref{fig:ResidCorrLX} not only support this interpretation of the
data, but also appear to suggest that the intrinsic flattening of a
galaxy, rather than its degree of rotational support, is what really
drives the $L_X$ deficiency of fast rotators.
When considering the correlations betwenn $\epsilon_{\rm e}$ and
either $L_X/L_{X,{\rm discr}}$ or $L_X/L_{X,{\rm discr+diff}}$, which
are the most significant in Fig.~\ref{fig:ResidCorrLX}, we note that
three of the roundest galaxies (NGC~3193, NGC~3640 and NGC~4753), are
likely to be intrinsically flat objects that are viewed face-on, which
would move to the right of the central panels of
Fig.~\ref{fig:ResidCorrLX} if seen from a less improbable angle.
This is supported by their large values for $\lambda_{\rm Re}$, which
indicates that they are in fact rotationally-supported systems just
like all the other fast rotators with which they share the trends with
$\lambda_{\rm Re}$ that is observed in the left panels of
Fig.~\ref{fig:ResidCorrLX}.
Conversely, in these same panels, the only slow rotators with $L_X$
values consistent with just emission from LMXBs or that fall short of
the predicted total $L_X$ from LMXBs and hot gas sustained by the
thermalisation of the kinetic energy of stellar ejecta, NGC~4365 and
NGC~5322, are incidentally the flattest objects among the class of
slow rotators.
In fact, that these (fairly massive) objects appear to align
themselves with the rest of the fast rotators in the anti-correlations
between the apparent flattening $\epsilon_{\rm e}$ and either
$L_X/L_{X,{\rm discr}}$ or $L_X/L_{X,{\rm discr+diff}}$ further
suggests that it is their intrinsic shape, rather than their dynamical
state, which determines whether they are capable to retain a halo of
hot gas.
When such flat slow rotators are excluded from the Spearman rank
analysis the correlations involving $\lambda_{\rm Re}$ in
Fig.~\ref{fig:ResidCorrLX} become more significant, and the same holds
when excluding the face-on fast rotators from the correlations
involving $\epsilon_{\rm e}$.
Finally, we note that even a modest contamination from a central AGN
or the the intra-group or -cluster medium to the total X-ray in the
least massive objects could significantly impact on the total $L_X$
values for these object. In turn, this would generally weaken the
significance of any correlation between $\lambda_{\rm Re}$ or
$\epsilon_{\rm e}$ and the $L_X/L_{X,{\rm discr}}$ or $L_X/L_{X,{\rm
    discr+diff}}$ ratios, since low-mass galaxies generally tend to be
fast rotators (Paper~III).
The possible impact of AGNs or the intra-group or -cluster medium is
most noticeable in the correlation between $\lambda_{\rm Re}$ and the
$L_X/L_{X,{\rm discr+diff}}$ deficit (lower left panel of
Fig.~\ref{fig:ResidCorrLX}), where the majority of the most
rotationally-supported galaxies with $L_X/L_{X,{\rm discr+diff}}$
values near unity have $\log{L_K\sigma_{\rm e}^2}$ values below of
$15.4 \,\rm L_{K,\odot}\,km^2\,s^{-2}$. 
At such a low $\log{L_K\sigma_{\rm e}^2}$ regime the typical AGN X-ray
emission $L_{X,{\rm AGN}}$ of early-type galaxies could be comparable
to the predicted values for $L_{X,{\rm discr}}$ and $L_{X,{\rm diff}}$
(Fig.~\ref{fig:LKsigLX}). For instance, the median $L_{X,{\rm AGN}}$
value for the \atlas\ sample galaxies that appear in the study of
\citet[][42 objects]{Liu11} is $2.6\times10^{39} \rm erg \,s^{-1}$.

To summarize the results of this subsection, in
Fig.~\ref{fig:LKLXLKsigLXcol} we redraw the same $L_K-L_X$ and
$L_K\sigma_{\rm e}^2-L_X$ diagrams presented in Figs.~\ref{fig:LKLX}
and \ref{fig:LKsigLX}, but now with elliptical symbols coded by colour
and flattening according to the values of $\lambda_{\rm Re}$ and the
$\epsilon_{\rm e}$ of the galaxies they are representing.
This is done in order to convey more visually our preliminary
conclusion, based on our low X-ray resolution sample, that only round
slow rotators can sustain a halo of hot gas from the thermalisation of
the kinetic energy carried by their stellar-mass loss material whereas
fast rotators (and apparently also flat slow rotators) have
progressively fainter X-ray halos the flatter and more
rotationally-supported they appear.

\subsection{High X-ray Resolution Sample}
\label{subsec:resultsHigh}

The previous conclusions on the more limited ability of fast-rotating,
or even more generally just flat, early-type galaxies to retain their
halos of hot gas have to be taken with care given that they are based
on X-ray data of rather coarse spatial resolution where the
contribution from a central AGN, the intra-group or -cluster medium or
LMXBs to the total $L_X$ values can only be estimated, and which in
the case of LMXBs is known to be the subject of considerable
scatter.\footnote{
 If we consider that globular clusters (GCs) may be the birthplace of
 the LMXB population in galaxies \citep[e.g.][]{Sar00,Whi02,Zha12} and
 that in turn lenticular galaxies may have a lower specific frequency
 $S_{N}$ of GCs than elliptical galaxies \citep{Kun01a,Kun01b}, it may
 be even possible that a systematically smaller number of LMXBs in
 fast rotators contributes to their lower X-ray luminosity. Nearly all
 (94\%) S0 galaxies and most (66\%) Es belong in fact to the class of
 fast-rotating galaxies (Paper~III).}
It is therefore important to ask whether our results hold when such
factors can be properly isolated through the use of \Chandra\ or \XMM.

\placefigLKsigLXgas

Recently, \cite{Bor11} published one of the largest samples of
early-type galaxies with consistently derived $L_{X,{\rm gas}}$
measurements, which overlaps well with the \atlas\ sample and excludes
the dominant members of groups and clusters of galaxies.
For the objects that we have in common with the sample of Boroson et
al., Fig.~\ref{fig:LKsigLXgas} shows how their $L_K\sigma_{\rm e}^2$
and $L_{X,{\rm gas}}$ values compare to each other, thus allowing to
directly check whether the hot-gas luminosity corresponds to the input
rate of kinetic energy from stellar-mass loss.
The use of the same color coding and squashing of
Fig.~\ref{fig:LKLXLKsigLXcol} for our data points makes it easy to
appreciate in Fig.~\ref{fig:LKsigLXgas} that, except for one, all the
slow rotators in this \atlas\ sub-sample agree still remarkably well
with the same simple prediction for $L_{X,{\rm diff}} = L_{\sigma}$
(Fig.~\ref{fig:LKsigLX} and right panel in
Fig.~\ref{fig:LKLXLKsigLXcol}).
Fast rotators, on the other hand, display $L_{X,{\rm gas}}$ values
that fall systematically short of such $L_{X,{\rm diff}}$ expected
values, as does also NGC~4365, which is the flattest slow rotator in
the high X-ray resolution sample and was also a low-lier compared to
the rest of the slow rotators in Fig.~\ref{fig:LKLXLKsigLXcol} (for a
more detailed picture of this galaxy, see also \citealp{Dav01},
\citealp{Sta04} and \citealp{vdB08}).

\placefigResidCorrLXgas

If Fig.~\ref{fig:LKsigLXgas} already seem to supports the picture that
was drawn in \S\ref{subsec:resultsLow} using \ROSAT\ or {\it
  Einstein\/} data, Fig.~\ref{fig:ResidCorrLXgas} further confirms
that the X-ray deficiency of fast rotators, as quantified by the
$L_{X,{\rm gas}}/L_{X,{\rm diff}}$ ratio, correlates indeed with their
degree of rotational support traced by the $\lambda_{\rm Re}$
parameter and apparent flattening $\epsilon_{\rm e}$.
The connection with $\lambda_{\rm Re}$ in particular is remarkably
significant given the modest number of objects at our disposal, and
the general trend in the left panel of Fig.~\ref{fig:ResidCorrLXgas}
can be recognised also in the lower-left panel of
Fig.~\ref{fig:ResidCorrLX} that showed the correlation between
$\lambda_{\rm Re}$ and the $L_{X}/L_{X,{\rm discr+diff}}$ deficit for
our low X-ray resolution sample, if one were to disregard (with the
benefit of hindsight) the least massive fast rotators.
The behaviour of the outliers to the present correlations with
$\lambda_{\rm Re}$ and $\epsilon_{\rm e}$ is also similar to what is
observed in Fig.~\ref{fig:ResidCorrLX}, with the flat slow rotator
NGC~4365 falling below the main trend between $L_{X,{\rm
    gas}}/L_{X,{\rm diff}}$ and $\lambda_{\rm Re}$ but aligning itself
with the majority of the fast rotators in the $L_{X,{\rm
    gas}}/L_{X,{\rm diff}}$ vs. $\epsilon_{\rm e}$ diagram and with a
few round fast rotators (NGC~3384, NGC~4278 and NGC~3379) acting in
the opposite way, which adds to the idea that the ability of a galaxy
to retain a halo of hot gas relates primarily to its intrinsic shape
\citep{Cio96,DEr98}.
In fact, that the correlation between the $L_{X,{\rm gas}}/L_{X,{\rm
    diff}}$ ratio and $\lambda_{\rm Re}$ is more significant
than in the case of the correlation with $\epsilon_{\rm e}$ suggests
that $\lambda_{\rm Re}$ may be a good indicator for the intrinsic
flattening of fast rotators, if this is indeed the main parameter
driving the X-ray deficit.

It is interesting to observe in the right panel of
Fig.~\ref{fig:ResidCorrLXgas} how the direct $L_{X,{\rm gas}}$
measurements derived from the \Chandra\ data now reveal that the
dynamical mass $M_{\rm dyn}$ must also play a role in determining how
much hot gas a galaxy can retain.
This is not unexpected, since the ratio of the energy that could be
injected by SNIa in the interstellar medium $L_{\rm SN}$ and the
energy required to steadily extract the mass lost by stars $L^-_{\rm
  grav}$ in itself depends on the galaxy mass, so that the less
massive galaxies would host SN-driven outflows and thus possess
systematically less massive and X-ray fainter hot-gas atmospheres.
In principle, it may be even possible that the correlation with
between $\lambda_{\rm Re}$ and $L_{X,{\rm gas}}$ is partly driven by
the correlation with $M_{\rm dyn}$, since low rotators tend to be more
massive then fast rotators (Paper~III).
Yet, a large $M_{\rm dyn}$ appears to be only a necessary, but not
sufficient condition for having a bright X-ray halo.
Almost all fast rotators in our high X-ray resolution sample are
nearly as massive, within a factor two, as the slow rotator NGC~4552
and yet they all display considerably smaller $L_{X,{\rm gas}}$ values.
Conversely, the flat slow rotator NGC~4365, shares with the flat
rotators of our sample a relatively faint hot gas halo despite being
more massive, by up to an order of magnitude, than nearly all of them.

We should also stress that the $L_{X,{\rm gas}}/L_{X,{\rm diff}}$
deficiency does not correlate with any of the indicators for the
environmental galactic density that were adopted in Paper~VII, which
also confirms that the impact the intra-cluster or group medium in our
high-resolution X-ray sample has been properly avoided by following
the selection of Boroson et al.

\citet{Bor11} also provide estimates for the hot gas temperature $T$,
which we can use to further test whether stellar motions are the main
heat source for the X-ray emitting gas \citep[see, e.g.,][]{Pel11}.
If this is the case, the gas temperature should relate to the stellar
velocity dispersion since this traces the stellar kinetic energy.
Traditionally $kT$ has been compared to $\mu m_p \sigma^2$ where $\mu
m_p$ is the mean particle mass of the gas ($m_p$ is the proton mass
and $\mu=0.62$ for Solar abundance) and $\sigma$ was taken as the
central stellar velocity dispersion, although $T$ should relate more
strictly to the stellar-density weighted average of the intrinsic,
three-dimensional velocity dispersion across the entire galaxy.
Rather than venturing directly in the construction of the stellar
dynamical models needed to properly evaluate this quantity, in
Fig.~\ref{fig:kTandAges} we plot the $kT$ values from Boroson et
al. against our integral-field measurements for the stellar velocity
dispersion within one effective radius $\sigma_{\rm e}$, which should
already fare better than the central $\sigma$ in tracing the global
stellar kinetic energy.\footnote{
 \citet{Pel11} shows that the kinetic temperature associated to the
 stellar motions can be expressed more generally as $T_{\star} = \mu
 m_p \sigma^2\Omega/k$, where $\Omega$ relates to the dark-matter
 fraction and to the shape of both the stellar and dark-matter density
 radial profiles. $\Omega$ is always $<1$ and typically range between
 $\sim0.6$ and $\sim0.8$ (see Fig.~2 of \citealp{Pel11}). This would
 include the value that the $\sigma_{\rm e}^2/\sigma^2$ ratio would
 take if the central $\sigma$ is measured within one eighth of the
 effective radius $R_e$ \citep[that is, 0.76, according to eq. (1)
   of][]{Cap06}, as is often the case in the literature.  $\mu m_p
 \sigma_e^2/k$ is therefore a good approximation for $T_{\star}$, just
 as it can be shown to be for $\frac{3}{2}\dot M \sigma_{\rm e}^2$ in
 the case of $L_{\sigma}$.}
In fact, by adopting the ${\rm \lambda_{\rm Re} =
  0.31\sqrt{\epsilon_{\rm e}}}$ dividing line between fast and slow
rotators introduced in Paper~III, we note in Fig.~\ref{fig:kTandAges}
that the hot-gas temperature of the slow rotators in our high X-ray
resolution sub-sample correlates well with $\sigma_{\rm e}^2$, with
$kT$ values that are on average above $\mu m_p \sigma_{\rm e}^2$ by
50\%.
On the other hand this does not seem to be the case for fast rotators,
which, with the exception of the particularly cold and under-luminous
X-ray halo of NGC~821, show similar $kT$ values between 0.25 and 0.35
keV across a range of $\sigma_{\rm e}^2$ values and where $\mu m_p
\sigma_{\rm e}^2$ seems to constitute just a lower limit for the
temperature of the hot gas.

\placefigeight

As already pointed out by \citet{Pel11}, the $kT$ excess of slow
rotators could be interpreted as a need for an extra source of
heating, possibly due to inflowing and adiabaticaly compressed gas in
the central regions of these objects, which would be consistent with
this $kT$ excess being itself proportional to $\sigma_{\rm e}^2$.
Yet, adding an extra heat source for the gas may not seem strictly
necessary given that the thermalisation of the stellar kinetic energy
can already account for the observed X-ray luminosity (see
Fig.\ref{fig:LKsigLXgas} and the right panel of
Fig.~\ref{fig:kTandAges}).
In fact, the agreement between the observed $L_X$ values and those
traced by $L_K\sigma_{\rm e}^2$ could be easily maintained even if we
were to scale up the kinetic energy traced by $\sigma_{\rm e}^2$ by
simply considering a smaller value for the stellar-mass rate $\dot M$,
as indicated for instance by the observations of \citet{Ath02} who
give $\dot M = 0.8 \times 10^{-11} (L_B/L_{\odot}) M_{\odot} \rm
yr^{-1}$ rather than the rate of $\dot M = 1.5 \times 10^{-11}
(L_B/L_{\odot}) M_{\odot} \rm yr^{-1}$ that we currently adopt from
\citet{Cio91}.

Why fast rotators do not follow the same $\sigma_{\rm e}^2 - kT$ relation of
slow rotators and instead scatter in the 0.25 -- 0.35 keV interval is
more difficult to understand.
One possibility is that in fast rotators $\sigma_{\rm e}^2$ may not
trace particularly well the intrinsic luminosity-weighted velocity
dispersion that would relate to the relative velocity involved in
those shocks between stellar ejecta that are supposed to thermalised
the kinetic energy inherited the stellar-mass loss material, in
particular since $\sigma_{\rm e}^2$ measures the total kinematic
broadening within one $R_{\rm e}$ and thus factors in also the impact
of net rotation.
On the other hand, the hot gas of fast rotators may be in an
altogether different dynamical state compared to slow rotators, being
most likely outflowing due to their flattening \citep{Cio96} or even
driven out in a supersonic wind powered by an additional number of
supernovae associated with a recently formed population of stars.
If gravitational heating associated to inflows is the reason why slow
rotators lie above the $kT = \mu m_p \sigma_{\rm e}^2$ line, then the
possible absence of inflows in fast rotators may explain why a good
fraction of them appear to fall close to this limit.
On the other hand, the color-coding of the symbols in
Fig.~\ref{fig:kTandAges} according to their global age (within one
$R_{\rm e}$, from McDermid et al. in preparation) points to an extra
input of energy from supernovae as the reason why another good
fraction of fast rotators display $kT$ values well above what is
expected from their $\sigma_{\rm e}^2$ (by a factor two), since most
of these objects show the presence of young populations.
The right panel of Fig.~\ref{fig:kTandAges} shows that such younger
objects seem also to show brighter X-ray halos compared to older fast
rotators, although more low-mass systems would certainly be needed to
confirm this impression.

Finally, we should also contemplate the possibility that the kinetic
energy inherited by the stellar mass-loss material is not thermalised
with the same efficiency in slow and fast rotators, in particular if
we consider that this process is supposed to happen through shocks as
the ejecta collide with each other or interact with an already
existing hot medium.
For instance, compared to the case of a dynamically supported system,
the stellar ejecta may find it harder to find each other in a galaxy
with net rotation, and eventually collide only at small angles and
relative velocities.
Furthermore, if we consider that the hot medium of a rotating galaxy
may have inherited part of the angular momentum of the stars from
which it originated\footnote{
 This is in fact quite possible given that the hot and warm phases are
 most likely in pressure equilibrium \citep[as the inferred values for
   the ionised-gas density suggest, see][and references
   therein]{Mat03} and considering that fast rotators can display
 extended ionised-gas disks that rotate in the same sense as the
 stars.
}, then at a given orbital velocity the newly shed ejecta of a
rotating galaxy may shock with the hot gas on average at a smaller
velocity than in the case of the ejecta of a non-rotating galaxy,
which would have imprinted no net rotation on its hot-gas halo.
According to these qualitative arguments we should expect that the
kinetic energy of stellar ejecta is thermalised less efficiently in
fast rotators.
\citet{Cio96} quantified this effect in the limit where none of the
stellar rotational motion is thermalised, finding that the hot-gas
temperature could decrease by up to 30\%.
Incidentally, in Fig.~\ref{fig:kTandAges} this seems to be fairly
close to the limit by how much colder the hot gas of fast rotators can
get compared to the hot gas of slow rotators.

To conclude we note that in Fig.~\ref{fig:kTandAges} the flat and
X-ray underluminous slow rotator NGC~4365 displays a $kT$ value in
line with the behaviour of the other members of its class.
This further indicates that if on one hand flattening could hamper the
ability of a galaxy to retain its hot gas, on the other hand the
degree of rotational support of a galaxy may also reduce the
efficiency with which the kinetic energy inherited by the stellar
ejecta is thermalised in the hot gas, which may be the case of fast
rotators but not of a dynamically hot system such as NGC~4365.

\subsection{Additional High X-ray Resolution Data}
\label{subsec:resultsMore}

If the analysis presented in the previous two sections suggests a
different specific hot-gas content of fast and slow rotators, we need
to keep in mind that these two kinds of early-type galaxies are rather
relegated to low and high values for their dynamical mass $M_{\rm
  dyn}$, respectively (Paper~III).
In fact, this is in particular the case of the objects in the high
X-ray resolution \atlas\ sub-sample that revealed also a possible role
of $M_{\rm dyn}$ in driving the X-ray deficit of early-type galaxies
(Fig.~\ref{fig:ResidCorrLXgas}, right panel), with the most massive
fast rotators showing $L_{X,{\rm gas}}$ values that appear nearly as
consistent with the input rate of kinetic energy from stellar-mass
loss as in the case of slow rotators (Fig.\ref{fig:LKsigLXgas} and the
right panel of Fig.~Fig.~\ref{fig:kTandAges}, where $M_{\rm dyn}$ is
fairly traced by $L_K\sigma_{\rm e}^2$).
It would be therefore useful to observe the behaviour of other
relatively massive fast rotators with $L_{X,{\rm gas}}$ estimates, as
well as of slow rotators with lower $M_{\rm dyn}$ values than
presently found in our high X-ray resolution sample, even if this will
entail a loss of consistency in the X-ray or optical measurements.

\placefignine

For this reason, in Fig.~\ref{fig:moreLKsigLXgas} we added to the
$L_K\sigma^2$ -- $L_{X,{\rm gas}}$ diagram also the other galaxies in
the Boroson et al. sample, using their adopted central values for
$\sigma$ as well as their $L_K$ and $L_{X,{\rm gas}}$ values, together
with four more objects in the \atlas\ sample with $L_{X,{\rm gas}}$
measurements from other relatively large studies based on
\Chandra\ data, which have a few objects in common with the Boroson et
al. that allow for some control on the consistency between $L_{X,{\rm
    gas}}$ estimates.
These \atlas\ objects, for which we can rescale the $L_{X,{\rm gas}}$
values using our adopted distance estimates and have access to the
rest of our \atlas\ imaging and integral-field spectroscopic
measurements, are the slow rotator NGC~6703 from the field sample of
\citet{Mul10} and the fast rotators NGC~4459, NGC~4494 and NGC~5845
from \citet{Ath07}.
Other $L_{X,{\rm gas}}$ compilations based on \Chandra\ data, such as
those of \citet{Die07} or \citet{Sun07} either did not include other
\atlas\ objects than those already considered by Boroson et al., or
only featured \atlas\ objects that are deeply embedded in their
intra-cluster or group medium.
Based on the deviations observed for the Boroson et al. objects that
are in the \atlas\ sample (and have thus been observed with \sauron)
we expect that for their remaining objects the $L_{X,{\rm gas}}$ and
$L_K\sigma^2$ measurements would be within 0.05 and 0.12 dex,
respectively, of the values that they would have if they had been
selected as an \atlas\ target, with no significant systematic
deviations.
\citeauthor{Mul10} have only two objects in common with the Boroson et
al. sample, with $L_{X,{\rm gas}}$ values that are consistent with
theirs (assuming the same distance).
\citet{Ath07} has 12 objects in common with Boroson et al., but in
this case the reported $L_{X,{\rm gas}}$ values are systematically
lower by 0.18 dex, albeit with a similar scatter.
Although these deviations are modest, for the sake of consistency, we
did not consider objects that are not in the \atlas\ sample with
$L_{X,{\rm gas}}$ other than those of Boroson et al.

To assign the Boroson et al. galaxies that are not part of the
\atlas\ sample to the slow- or fast-rotator families, we still adopted
the $\lambda_{\rm Re} = 0.31\sqrt{\epsilon_{\rm e}}$ dividing line of
Paper~III but now resort to an indirect estimate for $\lambda_{\rm
  Re}$ and adopt the 2MASS K-band measurement of the minor- to
major-axis ratio {\tt k\_ba} as a measure of the global flattening
$\epsilon_{\rm e}$.
To estimate $\lambda_{\rm Re}$ we use the relation between
$\lambda_{\rm Re}$ and the $(V/\sigma)_{\rm e}$ ratio given in
Paper~III (eq. B1) and in turn derive the $(V/\sigma)_{\rm e}$ ratio
by applying the correction of \citet[][their eq. 23]{Cap07} to our
best estimate of the $(V_{max}/\sigma)$ ratio from long-slit data in
the literature.
The only exception to this procedure is represented by NGC~720, which
is one of the ``special'' objects observed in the course of the
\sauron\ survey \citep{deZ02} and for which \citeauthor{Cap07} already
reported values for $\sigma_{\rm e}$, $\lambda_{\rm Re}$ and
$\epsilon_{\rm e}$. Based solely on its $\lambda_{\rm Re}$ value,
NGC~720 was considered a fast-rotator by \citet{Ems07}, albeit already
as an exceptional one by \citeauthor{Cap07}, and falls now in the
slow-rotator family when $\lambda_{\rm Re}$ is compared to
$0.31\sqrt{\epsilon_{\rm e}}$.

In Fig.~\ref{fig:moreLKsigLXgas} the objects with \atlas\ data (from
\citealp{Mul10} and \citealp{Ath07}) are colour-coded as in
Fig.~\ref{fig:kTandAges} according to our global age estimates (from
McDermid et al. in preparation) whereas for the remaining galaxies of
the Boroson et al. sample we limited ourselves to identify only
objects with clear signs of recent star-formation episodes.
These are NGC~1316, for which \citet{Sil08} used near-IR data to
establish a mean stellar age of $\sim\!2$ Gyr, and NGC~1052, for which
\citet{Fer11} report a global value for the NUV-r colour of 4.9,
typical for early-type galaxies that experienced a recent
star-formation episode \citep[e.g.][]{Kav07}.
For the remaining sample galaxies of Boroson et al. it is difficult to
draw a consistent picture of their stellar age, given the varying
quality of the absorption-line strength measurements found in the
literature (not always corrected for nebular emission) and the
different choices for the models used to derive the stellar population
properties.

The inclusion of these additional objects in our analysis confirms
that fast rotators tend to have fainter X-ray halos than slow rotators
with similar $L_{X,{\rm diff}}$ expectation values (from the
thermalisation of the kinetic energy brought by the stellar-mass loss
material), although Fig.~\ref{fig:moreLKsigLXgas} further indicates
that such a discrepancy may reduce or even disappear as the dynamical
mass of galaxies increases.
As in the previous exceptional cases of NGC~4365 and NGC~5322, the
slow rotator with the relatively faintest hot-gas halo among the
additional objects, NGC~1427, also happens to be a relatively flat
galaxy ({\tt k\_ba}=0.7; see \citealp{DOn95} for its long-slit
kinematics).
On the other hand, Fig.~\ref{fig:moreLKsigLXgas} introduces in our
picture two flat and well-studied slow rotators, NGC~720 and NGC~3923
\citep[with a {\tt k\_ba} value of 0.55 and 0.64, respectively; for
  the kinematics of NGC~3923 see][]{Pel97}, for which $L_{X,{\rm
    gas}}$ appears well in line with the $L_{X,{\rm diff}}$
prediction.
Thus, it would appear that flattening can, but not necessarily does,
drive a deficit in hot-gas content of slow rotators.
Finally, younger stellar populations are found in three out of the
four fast rotators that in Fig.~10 sit the closest to the $L_{X,{\rm
    diff}}$ prediction, consistent with a possible extra energy input
from SNe in younger objects.

\section{Discussion}
\label{sec:discussion}

By combining photometric and integral-field spectroscopic measurements
with X-ray data of both low and high spatial resolution we have found
that flat and rotationally-supported early-type galaxies (i.e., fast
rotators) tend to fall systematically short of the X-ray luminosity
$L_{X,{\rm gas}}$ that is expected if the hot gas radiation is
sustained by the thermalisation of the kinetic energy $L_{\sigma}$
that the stellar-mass loss material inherits from their parent stars.
On the other hand, non- or slowly-rotating rotating galaxies (i.e.,
slow rotators), which are likely to be also intrinsically fairly round
(Weijmans et al., in preparation), generally display $L_{X,{\rm gas}}$
values in line with this model, and in fact also show hot-gas
temperatures that correspond well to the stellar kinetic energy, as
traced by the global stellar velocity dispersions within the optical
regions of these galaxies $\sigma_{\rm e}$.
This is not the case for fast rotators, which show nearly the same
hot-gas temperature across a range of $\sigma_{\rm e}$ values except
for a few objects with younger stellar population that would appear to
have hotter and also brighter gas halos, possibly owing to the
additional energy input from more recent supernovae explosions.
Such a discrepancy in the hot-gas content of fast and slow rotators
appears to reduce, or even disappear, for large values of the
dynamical mass (beyond $\sim3\times10^{11}M_{\odot}$, corresponding in
the \atlas\ sample to $\log{L_K\sigma_{\rm e}^2}\sim16$), whereas the
few slow rotators with $L_{X,{\rm gas}}$ values considerably below the
bar set by the thermalisation of the kinetic energy brought by
stellar-mass loss material always appear to be relatively flat.
This last result could suggest that it is the intrinsic shape of
galaxies, rather than their degree of rotational support, that reduces
their ability to retain the hot gas \citep{Cio96}, although care is
needed as there exist also flat slow rotators (albeit very massive)
with $L_{X,{\rm gas}}$ perfectly in line with the $L_{X,{\rm diff}}$
expectation values.

The situation observed for slow-rotators, whereby $L_{X,{\rm gas}}
\sim L_{\sigma}$, is likely due to the fact that currently $L_{\rm SN}
\sim L_{\rm grav}^-$ in these galaxies, so that most the SNe type Ia
energy is presumably used to steadily remove the stellar-mass loss
material and thus maintain a quasi-static hot-gas
atmosphere.\footnote{
 In fact, as discussed in \citet{Pel98} and \citet{Pel12}, even in
 such a quasi-static situation the gas flow is likely to be decoupled
 into an outflowing outskirt and a central inflow where radiative
 cooling outstrips the SNe energy feedback.}
On the other hand, by virtue of their flattening and possibly also of
their lower mass, $L_{\rm SN}$ exceeds $L_{\rm grav}^-$ in fast
rotators so that these systems are actually degassing and thus
display more tenuous and X-ray fainter hot-gas halos.

A systematic offset in the specific hot-gas content of fast and slow
rotators may be relevant to understand also the disparity between the
incidence of molecular gas \citep{You11} and associated episodes of
recent star formation \citep{Kun10} in these two classes of early-type
galaxies (with slow rotators being devoid of both molecular gas and
young stars), which in turn may link to an altogether different origin
and fate for their gas \citep[][]{Sar07}.
Indeed, much as the intra-cluster or group medium may suppress the
cold-gas reservoir of galaxies falling in crowded galactic enviroments
and drive the cosmic rise of lenticular galaxies
\citep[e.g.][]{Dre97}, the presence of a halo of hot gas around a
galaxy may inhibit the star formation triggered either by the
acquisition of gas-rich satellites or by the accumulation of
stellar-mass loss material.
This is because small amounts of fresh gas would quickly heat up and
evaporate in such an hot medium, by thermalising its kinetic energy as
it moves against the hot gas and inevitably shocks with it
\citep[e.g.][for stellar winds]{Par08}, and possibly also through
thermal conduction \citep[e.g.][for accreted gas]{Nip07}.
Furthermore, if hot-gas halos inherit part of the angular momentum of
the galaxies that they enshroud (through the interaction between the
hot gas and the stellar-mass loss material, as advocated by
\citealp{Mar11} in the case of disk galaxies), the accretion of gas
from satellites could proceed differently depending on whether such
material travels with or against the hot gas.
In the latter case, for instance, ram-pressure stripping would be more
efficient, since its strength depends on the square of the velocity
relative to the hot medium.

Based on the distribution of the misalignment between the angular
momentum of stars and ionised gas, \citet{Sar06} already recognised
(based on the smaller \sauron\ representative sample) a possible
difference in the relative role of external gas accretion and
stellar-mass loss in fast and slow rotators, where in the latter class
the warm gas would appear purely external in origin.
Using the \atlas\ sample, \citet{Dav11} not only confirmed this
difference but also highlighted the impact of both a dense galactic
environment (that is, inside Virgo) and a large mass (for $M_K < -24$
or $\log{L_K}=10.9$) in suppressing the acquisition of external gas by
fast rotators.
A bias towards finding corotating gas and stars in the most massive
fast rotators is also consistent with the conclusion of \citet{Sha10}
that there exists two modes of star-formation in fast rotators, such
that in low-mass systems star-formation is found throughout the galaxy
and is triggered by minor mergers whereas in galaxies more massive
that $\log{M_{\rm dyn}}\sim10.95$ (which in the \atlas\ sample
corresponds to $\log{L_K}\sim10.8$) such an activity occurs only in
central or circumnuclear regions and the ionised-gas rotates in the
same sense as the stars.
Interestingly, slow rotators would appear to accrete their ionised-gas
despite tending to be more massive and to reside in denser galactic
environments than fast rotators \citep{Dav11}.
Perhaps, the present tendency of slow rotators to experience minor
mergers can be regarded as a snapshot of their overall assembly
history, since numerical simulations indicate that such a constant
``bombardment'' is required in order to reduce the specific stellar
angular momentum to the values observed in this class of objects
\citep[][hereafter Paper~VI and VIII, respectively]{Boi11,Kho11}.

In any event, the gas acquired in slow rotators does not appear to
condense to a molecular phase and form new stars.
Instead it must join the hot interstellar medium, most likely through
thermal conduction as first suggested by \citet{Spa89} and later
quantified by \citet{Nip07}.\footnote{
 As a reference, \citeauthor{Nip07} find that gas clouds of mass below
 $10^4-10^5\rm M_{\odot}$ and $10^7-10^8 \rm M_{\odot}$ (at most, they
 are more likely to be 100 times smaller) would eventually evaporate
 as they fall in the hot medium of galaxies of baryonic mass of
 $3\times10^{10}\rm M_{\odot}$ and $3\times10^{11}\rm M_{\odot}$,
 respectively.
}
In fact, the lack of slow rotators with corotating gas and stars means
that a similar fate also awaits any stellar-mass loss material
injected in such hot-gas rich systems, whereas the systematic
$L_{X,{\rm gas}}$ deficiency of fast rotators may allow for some
fraction of this material to cool down amidst their more tenuous hot
medium.

A good way of testing whether fast rotators, as opposed to slow
rotators, can manage to cool down a fraction of their stellar
mass-loss material - thus possibly being more efficient at recycling
such gas into new stars - would be to compare the specific content of
the diffuse dust in these two classes of objects.
Observed at infrared wavelengths, such diffuse dust can indeed trace
more directly the material lost by stars than done for instance by the
ionised-gas component of the interstellar medium, in particular since
the nebular flux may reflect also a varying UV ionising flux from a
number of different sources \citep{Sar10} whereas the dust of
quiescent galaxies is mainly heated by the optical stellar radiation
\citep[e.g.][]{Sau92}.
Furthermore, the launch of the {\it Herschel\/} Space Observatory has
now made it possible to probe the colder dust component of early-type
galaxies and thus estimate more accurately their total dust mass
$M_{\rm dust}$ than previously allowed by less sensitive instruments
covering a more limited wavelength range, such as {\it IRAS\/}, {\it
  ISO\/} or {\it Spitzer\/} \citep[see,
  e.g.,][]{Gou95,Fer02,Ath02,Tem07}.

Recently \citet{Smi12} published $M_{\rm dust}$ estimates based on
{\it Herschel\/} data for a sample of 62 nearby early-type galaxies,
which include 57 objects from the \atlas\ sample.
Unfortunately, within this \atlas\ sub-sample there are only five
galaxies with both $M_{\rm dust}$ secure detections and $L_{X,{\rm
    gas}}$ measurements, making a direct comparison between the
specific dust content and the X-ray properties in our sample galaxies
not very meaningful.
On the other hand, since the \atlas\ objects with $M_{\rm dust}$
estimates have a range of $M_{\rm dyn}$ and $L_K\sigma_{\rm e}^2$
values that matches well that of the high X-ray resolution
\atlas\ sub-sample and of the additional objects that we have analysed
in \S~\ref{subsec:resultsHigh} and \S~\ref{subsec:resultsMore},
respectively, we can use our results to gauge the $L_{X,{\rm gas}}$
deficiency of the \citeauthor{Smi12} galaxies in our sample, and
correlate that with their specific dust content.

\placefigten

For this reason, in Fig.~\ref{fig:Herschel} we show how the ${\rm
  \lambda_{\rm Re}/(0.31\sqrt{\epsilon_{\rm e}}})$ ratio correlates
with the specific dust content $M_{\rm dust}/L_K$ of the
\atlas\ objects analysed by \citeauthor{Smi12}
This is similar to Fig.~12 of that work, except that here we use $L_K$
as a measure of the total stellar mass and regard ${\rm \lambda_{\rm
    Re}/(0.31\sqrt{\epsilon_{\rm e}}})$ as an optical tracer of the
$L_{X,{\rm gas}}$ deficiency compared to what is expected from the
thermalisation of the kinetic energy of the stellar-mass loss material
(\S~\ref{subsec:resultsHigh}).
Furthermore, in Fig.~\ref{fig:Herschel} we also show objects with only
upper-limits on $M_{\rm dust}$ and color-code those with secure
$M_{\rm dust}$ detections according to whether their ionised-gas
component shows an angular momentum that is considerably misaligned
compared to that of the stars (i.e. by more than 30$^{\circ}$).

Fig.~\ref{fig:Herschel} shows that fast rotators generally extend to
larger values for the specific dust content than slow rotators, with
very few fast rotators showing $M_{\rm dust}/L_K$ values reaching down
to the values observed in the more dust-poor slow rotators.
A similar message is conveyed also by the upper-limits on $M_{\rm
  dust}/L_K$, in that they tend to constrain the specific dust content
to consistently lower values as ${\rm \lambda_{\rm
    Re}/(0.31\sqrt{\epsilon_{\rm e}}})$ decreases.
In fact, all the slow rotators with secure $M_{\rm dust}$ detections
in Fig.~\ref{fig:Herschel} show largely decoupled nebular and stellar
motions consistent with an external origin for their gas, which
suggests that the specific content of internally generated dust in
these systems is probably much lower than currently inferred from
their total far-infrared emission.
On the other hand, most of the fast rotators with a specific dust
content in excess of what is observed in slow rotators (that is for
$M_{\rm dust}/L_K > -4.5$) display well-aligned gas and stellar
angular momenta (in nine out of ten cases, corresponding at most to a
20\% fraction of systems where the gas was accreted), suggesting that
such a larger dust content originates from stellar-mass loss material.
This finding does not appear to be a result of sample selection. For
instance, taken together all the fast rotators with $M_{\rm dust}$
detections are consistent with a even split between objects with
internal and external origin for their gas just like for the entire
\atlas\ population of fast rotators \citep{Dav11}.
Furthermore, the fraction of objects in high-density environments
(i.e., within the Virgo cluster) or with relatively large mass (i.e.,
$M_K < -24$) does not change when considering all the fast rotators
with $M_{\rm dust}$ detections or just those with a higher specific
dust mass content. This suggests that these otherwise important
factors in determining the origin of the gas cannot entirely explain
why the dustier fast rotators in Fig.~\ref{fig:Herschel} also happen
to have co-rotating gas and stars.

Thus, once complemented with our kinematic information, the {\it
  Herschel\/} data of \citeauthor{Smi12} would appear to confirm that
fast rotators have a larger specific dust content which relates to the
fact that some fraction of the stellar-mass loss material is allowed
to cool in the less hostile hot-gas medium of these objects, compared
to the case of slow rotators.
Such an ability of preserving some of their stellar-mass loss material
from joining the hot medium could also contribute (together with our
previous conjecture on the possible role of a rotating X-ray halo) to
the finding of \citet{Dav11} that the most massive fast rotators do
not seem to further acquire external gas, even when found in
relatively sparse galactic environments.
In fact, it may be even possible that in time such fast rotators build
up a gaseous disk that is sufficiently massive to absorb
counter-rotating gas from smaller satellites, which, on the other
hand, would perturb more significantly the structure of less massive
fast rotators.
Yet, it is important to keep in mind here that not all fast rotators
display dust or ionised-gas, pretty much independent of galaxy mass
and only mildly dependent on galactic environment \citep[see Tab.~2
  of][]{Dav11}.
Even low-mass systems in the field, where both external accretion and
internal recycling could occur, can be devoid of any gas. Apparently
other mechanisms, beyond the scope of this paper, must operate to also
remove gas from early-type galaxies.

A different specific hot-gas content of slow and fast rotators and the
likely different fate of their stellar-mass loss material could also
contribute to understand the fraction of rejuvinated early-type
galaxies that has been estimated in much larger spectroscopic surveys
of more distant galaxies.
For instance, based on a consistent treatment of Sloan Digital Sky
Survey spectra for over 3,000 morphologically selected early-type
galaxies, \citet{Tho10} revised the conclusion of \citet[][based on a
  more heterogeneous sample]{Tho05} and found that the impact of
environment on the fraction of rejuvinated galaxies decreases with the
mass of galaxies.
In practice, for values of $M_{\rm dyn}$ above $\sim 10^{10.75} \rm
M_{\odot}$ the fraction of galaxies with evidence of recent star
formation steadily decreases till reaching values consistent with zero
pretty much independent of galactic environment \citep[see Fig~8
  of][]{Tho10}.
Given that our results are based on galaxies exceeding similar values
of $M_{\rm dyn}$ (Fig.~~\ref{fig:ResidCorrLXgas}) we suggest that the
internal processes regulating the star-formation history at the
massive end of the early-type galaxy population could be driven by the
hot-gas content of these objects and their ability to recycle the
stellar-mass loss material into new stars, which in turn could relates
to the relative number of slow and fast rotators as well as the
different degrees of $L_{X,{\rm gas}}$ deficiency displayed by the
latter class of objects.
For instance, where the fraction of rejuvenated objects in
\citeauthor{Tho10} starts to be negligible, around $M_{\rm
  dyn}\sim10^{11.25} \rm M_{\odot}$, is incidentally also the point
from which the fraction of slow rotators in the \atlas\ sample
shoots-up rapidly from values around 5--15\% to a fraction of
$\sim$60\% for $M_{\rm dyn}\sim10^{11.5} \rm M_{\odot}$ (Fig.~11 of
Paper~III, but see also Fig.~10 of Cappellari et al. 2012c, hereafter
Paper~XX), by which mass the hot-gas content of fast rotators may not
longer differ too much from that of slow rotators.

To conclude, we note that gravitational heating of infalling gas
during minor mergers \citep[in particular around the most massive
  galaxies,][]{Dek06, Kho08, Jon09} is expected to deposit significant
amount of energy in the hot gas.
The relative contribution of this process compared to the
thermalization of the kinetic energy inherited by the stellar-mass
loss material remains to be quantified, but unless to consider a very
small efficiency for the latter mechanism, our results would suggest
that - at least when considering the hot gas around galaxies that is
not bounded by a group or cluster potential - gravitational heating
presently does not play an dominant role.

%
\section{Conclusions}
\label{sec:conc}

For a galaxy, being embedded in a corona of hot, X-ray emitting gas
can be a key element determining its recent star-formation history. A
halo of hot gas can indeed act as an effective shield against the
acquisition of cold gas and can quickly absorb any stellar-mass loss
material, thus preventing its cooling and consequent recycling into
new stars.
In other words, the ability to sustain a hot-gas halo can contribute
to keep a galaxy in the so-called red-sequence of quiescent objects.

Since the discovery by the {\it Einstein} observatory of such X-ray
halos around early-type galaxies, the origin of the X-ray emission and
the precise amount of hot gas around these galaxies have been the
matter of long debates, in particular when trying to understand the
rather loose correlation between the optical and X-ray luminosity of
early-type galaxies.
In the past, this situation resulted from the limited ability to
isolate in earlier X-ray data the additional contribution from an
active nucleus, the unresolved population of X-ray binaries and the
X-ray emission from the intra-cluster medium, although the use of
heterogeneous optical data may have also contributed to such an
impasse.
Today, with new X-ray telescopes such as \Chandra\ or \XMM\ and large
collections of both photometric and spectroscopic data it is possible
to gain more insight on the hot-gas content of early-type galaxies.

In this paper we have combined the homogeneously-derived photometric
and spectroscopic measurements for the early-type galaxies of the
\atlas\ integral-field spectroscopic survey with measurement of their
X-ray luminosity from X-ray data of both low and high spatial
resolution, finding that the hot-gas content of early-type galaxies
can depend crucially on their dynamical structure.
Specifically, in the framework of the revised kinematic classification
for early-type galaxies advanced in the course of both the
\sauron\ and \atlas\ surveys \citep[][Paper~III]{Ems07}, we find that:

\begin{itemize}

\item Slow rotators have hot-gas halos with X-ray luminosity values
  $L_{X,{\rm gas}}$ that are generally consistent - across a fair
  range of $M_{\rm dyn}$ values - with what is expected if the hot gas
  radiation is sustained by the thermalisation of the kinetic energy
  that the stellar-mass loss material inherits from their parent
  stars, so that $L_{X,{\rm gas}}$ closely follows $L_K\sigma_{\rm
    e}^2$.

\item Fast rotators display $L_{X,{\rm gas}}$ values that tend to fall
  short of the prediction of such a model, and the more so the larger
  their degree of rotational support as quantified using the
  $\lambda_{\rm R}$ integral-field parameter.

\item Such a $L_{X,{\rm gas}}$ deficiency in fast rotators would
  appear to reduce, or even disappear, for large values of the
  dynamical mass (beyond $\sim3\times10^{11}M_{\odot}$), whereas the
  few slow rotators with $L_{X,{\rm gas}}$ values considerably below
  the bar set by the thermalisation of the kinetic energy brought by
  stellar-mass loss material always appear to be relatively flat,
  nearly as much as on average fast rotators do.

\item Still consistent with a stellar origin for the heat needed to
  sustain their hot-gas emission, slow rotators also display hot-gas
  temperatures that correspond well to the global stellar kinetic
  energy, as traced by the stellar velocity dispersions measured
  within the optical regions $\sigma_{\rm e}$.

\item Fast rotators, on the other hand, show similar values of $T$
  across a range of $\sigma_{\rm e}$, except for a few objects with
  younger stellar population that would appear to have both hotter and
  brighter X-ray halos, possibly owing to the additional energy input
  from more recent supernovae explosions.

\end{itemize}

These results indicate that the thermalisation of the kinetic energy
brought by stellar-mass loss material, presumably through shocks and
collisions of the ejecta with each other and the hot interstellar
medium, sets a natural upper limit to X-ray emission originating from
the hot gas bounded by the gravitational potential of present-day
quiescent galaxies.
This is observed mostly in the case of slow rotators, where the energy
brought by SNe type Ia is similar to that needed to steadily extract
most of the mass lost by stars, thus helping maintain a quasi-static
hot-gas atmosphere.
On the other hand, the fact that fast rotators as a class are most
likely to be intrinsically flatter than slow rotators (with fast
rotators being nearly as flat as spiral galaxies and slow rotators
being much rounder and possibly triaxial, Weijmans et al. in
preparation) leads us to interpret the $L_{X,{\rm gas}}$ deficit of
fast rotators in light of the scenario of \citet{Cio96} whereby
flatter galaxies have a harder time in retaining their hot gas.
Flattening indeed reduces the binding energy of the gas, which would
cause SNe to drive global outflows in fast rotators, thus
reducing their hot-gas density and X-ray luminosity.
The finding that the few $L_{X,{\rm gas}}$-deficient slow rotators
also happen to be relatively flat would lend further support to this
hypothesis, although care is needed as there are also flat slow
rotators with $L_{X,{\rm gas}}$ perfectly in line with what is
observed in the more standard and round slow rotators.
We also note that if the intrinsic shape of a galaxy could determine
the ability of a galaxy to retain its hot-gas halo, the degree of
rotational support could further lower the efficiency with which the
kinetic energy carried by the stellar-mass loss material is
thermalised in the hot gas.
Indeed, this efficiency appears to reach its peak in the dynamically
supported slow rotators, where $T$ traces well $\sigma_{\rm e}$.

Finally, using recent {\it Herschel} measurements for the diffuse dust
emission of early-type galaxies and the kinematic information on the
relative motions of stars and gas provided by our integral-field
spectroscopic observations, we have also shown that fast rotators have
a larger specific dust content compared to slow rotators, and that
this is likely due to the fact that some fraction of the stellar-mass
loss material is allowed to cool down in the more tenuous medium of
these objects.
Such an ability to recycle the stellar-mass loss links well with the
fact that among massive early-type galaxies, only fast rotators have
detected molecular gas and display signs of recent star-formation,
something that cannot occur in slow rotators where both stellar-mass
loss and accreted material quickly fizzles in the hot medium.

\section*{Acknowledgements}
We wish to thank the anonymore referee, whose comments helped much in
improving the paper. MS is indebted to Gary Mamon, Silvia Pellegrini,
Nicola Brassington, Martin Hardcastle and Luca Cortese for useful
discussion, and to both ESO and the Institut d'Astrophysique de Paris
for their hospitality during much of the preparation of this paper. MS
acknowledges support from an STFC Advanced Fellowship ST/F009186/1. MC
acknowledges support from a Royal Society University Research
Fellowship. This work was supported by the rolling grants
`Astrophysics at Oxford' PP/E001114/1 and ST/H002456/1 and visitors
grants PPA/V/S/2002/00553, PP/E001564/1 and ST/H504862/1 from the UK
Research Councils. RLD acknowledges travel and computer grants from
Christ Church, Oxford and support from the Royal Society in the form
of a Wolfson Merit Award 502011.K502/jd. SK acknowledges support from
the Royal Society Joint Projects Grant JP0869822.  RMcD is supported
by the Gemini Observatory, which is operated by the Association of
Universities for Research in Astronomy, Inc., on behalf of the
international Gemini partnership of Argentina, Australia, Brazil,
Canada, Chile, the United Kingdom, and the United States of
America. TN and MBois acknowledge support from the DFG Cluster of
Excellence `Origin and Structure of the Universe'. NS and TD
acknowledge support from an STFC studentship. The authors acknowledge
financial support from ESO. We acknowledge use of the NASA/IPAC
Extragalactic Database (NED) which is operated by the Jet Propulsion
Laboratory, California Institute of Technology, under contract with
the National Aeronautics and Space Administration.

%

\label{lastpage}
\end{document}